%% file: main.tex
\documentclass[a4paper,fleqn]{cas-sc}

\input{preamble}
\begin{document}

\let\WriteBookmarks\relax
\def\floatpagepagefraction{1}
\def\textpagefraction{.001}

\shorttitle{PyFR v2.0.3}    
\shortauthors{}
\title[mode = title]{PyFR v2.0.3: Towards Industrial Adoption of Scale-Resolving Simulations}

% Large author file
\input{authors}

\begin{abstract}
PyFR is an open-source cross-platform computational fluid dynamics framework based on the high-order Flux Reconstruction approach, specifically designed for undertaking high-accuracy scale-resolving simulations in the vicinity of complex engineering geometries. Since the initial release of PyFR v0.1.0 in 2013, a range of new capabilities have been added to the framework, with a view to enabling industrial adoption of the capability. This paper provides details of those enhancements as released in PyFR v2.0.3, explains efforts to grow an engaged developer and user community, and provides latest performance and scaling results on up to 1024 AMD Instinct MI250X accelerators of Frontier at ORNL (each with two GCDs), and up to 2048 NVIDIA GH200 GPUs on Alps at CSCS.
\end{abstract}


\maketitle

% All CPiP articles must contain the following
% PROGRAM SUMMARY.

\noindent
{\bf PROGRAM SUMMARY}

\begin{small}
\noindent
{\em Program Title:} PyFR \\
{\em CPC Library link to program files:}  \\
{\em Developer's repository link:} https://github.com/PyFR/PyFR \\
{\em Code Ocean capsule:} \\
{\em Licensing provisions:} BSD 3-clause  \\
{\em Programming language:} Python (generating C/OpenMP, CUDA, OpenCL, HIP, Metal)                                   \\
{\em Nature of problem:}
  Accurate and efficient scale-resolving simulation of industrial flows.\\
{\em Solution method:}
  Massively parallel cross-platform implementation of high-order accurate Flux Reconstruction schemes.
\end{small}

\input{introduction}
\input{fr}

\input{additions}

\input{community}

\input{frontier}
\input{conclusions}
\input{acknowledgements}

\bibliographystyle{unsrtnat} 
\bibliography{reference}

\end{document}

%% file: preamble.tex
\usepackage{amssymb}
\usepackage{amsmath}
\usepackage{amsfonts}
\usepackage{upgreek}
\usepackage{anyfontsize}
\usepackage{mathtools}
\usepackage{siunitx}
\usepackage{listings}
\usepackage{multirow}
\usepackage{multicol}
\usepackage[sort&compress,square,numbers]{natbib}
\usepackage{siunitx}
\usepackage{color}
\usepackage{inconsolata}
\usepackage{soul}
\usepackage{booktabs}
\usepackage[section]{placeins} % For '\input{file}'
\usepackage[titletoc]{appendix}
\usepackage{calc}
\usepackage[capitalize]{cleveref}
\usepackage{tabularx}

\usepackage{xspace}

\usepackage{graphicx}
\usepackage{graphics}
\usepackage{float}
\usepackage{subfloat}
\usepackage{caption}
\usepackage{subcaption}
\usepackage{wrapfig}
\usepackage[percent]{overpic}
\usepackage{varwidth}
\usepackage{adjustbox}

\usepackage{tikz}
\usepackage{tikz-3dplot}
\usepackage{pgfplots}
\usetikzlibrary{arrows,matrix,positioning,fit}
\usetikzlibrary{shapes,positioning}
\usetikzlibrary{shapes.multipart}
\usetikzlibrary{backgrounds}
\pgfplotsset{compat=1.14}
\usepgfplotslibrary{colorbrewer}
\usepgfplotslibrary{patchplots}
\usepgfplotslibrary[colorbrewer]
\usetikzlibrary{pgfplots.colorbrewer}
\usetikzlibrary[pgfplots.colorbrewer]
\usepgfplotslibrary{fillbetween}
\usepackage{tikz-layers}
\usepgfplotslibrary{units}
\usetikzlibrary{spy}
\usepackage{pgfplotstable}
\graphicspath{ {figs/} }
\newlength\myheight
\newlength\mydepth
\settototalheight\myheight{Xygp}
\DeclareRobustCommand\onedot{\futurelet\@let@token\@onedot}
\def\@onedot{\ifx\@let@token.\else.\null\fi\xspace}

\makeatother

%% file: authors.tex

% Authors will appear in the order they are added
\author[tamu]{Freddie D. Witherden} % agreed

\author[icl]{Peter E. Vincent} % agreed
\cormark[1] % corresponding author mark
\ead{p.vincent@ic.ac.uk}

\author[ibm]{Will Trojak} % agreed
\author[toho]{Yoshiaki Abe} % agreed
\author[tamu]{Amir Akbarzadeh} % Asked, confirmed
\author[icl]{Semih Akkurt} % agreed
\author[tamu]{Mohammad Alhawwary}
\author[icl]{Lidia Caros} % agreed
\author[llnl]{Tarik Dzanic} 
\author[siem]{Giorgio Giangaspero} % agreed
\author[ast]{Arvind S. Iyer} % agreed
\author[tamu]{Antony Jameson} % Freddie
\author[niv]{Marius Koch} % agreed
\author[nivh]{Niki Loppi} % agreed
\author[tamu]{Sambit Mishra} % Asked
\author[tamu]{Rishit Modi} % Asked
\author[lum,isae]{Gonzalo Sáez-Mischlich}
\author[inh]{Jin Seok Park} % agreed
\author[con]{Brian C. Vermeire} % agreed
\author[unk]{Lai Wang} % Agreed, no affil

% Affiliations ##########################
% TODO - Finalise addresses. Peter

\affiliation[tamu]{organization={Department of Ocean Engineering, Texas A\&M University},
            addressline={3145 TAMU}, 
            city={College Station},
            postcode={77843}, 
            state={TX},
            country={USA}}

\affiliation[icl]{organization={Department of Aeronautics, Imperial College London},
            addressline={South Kensington}, 
            city={London},
            postcode={SW7 2AZ}, 
            country={UK}}
            
\affiliation[ibm]{organization={IBM Research UK},
            addressline={Hartree Centre}, 
            city={Warrington},
            postcode={WA4 4AD},
            country={UK}}

\affiliation[toho]{organization={Institute of Fluid Science, Tohoku University},
            addressline={Sendai},
            country={JP}}

\affiliation[llnl]{organization={Lawrence Livermore National Laboratory},
            addressline={Livermore}, 
            city={CA},
            country={USA}}

\affiliation[siem]{organization={Siemens Digital Industries Software},
            addressline={Shepherds Bush Road},
            city={London},
            postcode={W6 7NL},
            country={UK}}

\affiliation[ast]{organization={Astrome Technologies},
            addressline={Bangalore}, 
            country={IN}}

\affiliation[niv]{organization={NVIDIA Corporation},
            addressline={Fasanenstr 81},
            city={Berlin},
            postcode={10623},
            country={DE}}
            
\affiliation[nivh]{organization={NVIDIA Corporation},
            addressline={Porkkalankatu 1},
            city={Helsinki},
            postcode={00180},
            country={FI}}
            
\affiliation[lum]{organization={Luminary},
            addressline={101 S Ellsworth Ave Suite 600}, 
            city={San Mateo},
            state={CA},
            country={USA}}

\affiliation[isae]{organization={ISAE SUPAERO}, addressline={University of Toulouse}, city={31400 Toulouse}, country={FR}}
            
\affiliation[inh]{organization={Department of Aerospace Engineering, Inha University},
            addressline={Incheon},
            country={KR}}

\affiliation[con]{organization={Department of Mechanical, Industrial, and Aerospace Engineering, Concordia University},
            addressline={Montreal}, 
            city={QC},
            country={CA}}

\affiliation[unk]{organization={Gaithersburg},
            city={MD},
            country={USA}}

%% file: introduction.tex
\section{Introduction}

Computational Fluid Dynamics (CFD) is used by high-value industries across the world to reduce costs and improve product performance. The majority of industrial CFD is undertaken using Reynolds-Averaged Navier--Stokes (RANS) simulations, which time-average unsteady phenomena, including turbulence, and replace the `missing physics’ with a model. However, it is well established that RANS approaches have limited applicability when flow is separated and unsteady. To overcome this limitation, higher-fidelity scale-resolving methods can be used, such as Large Eddy Simulations (LES), Implicit Large Eddy Simulations (ILES) and Direct Numerical Simulations (DNS), with DNS being the most accurate; fully resolving all physics of the governing Navier--Stokes equations. However, the cost of a scale-resolving simulation is typically orders-of-magnitude higher than that of a RANS simulation, and thus the use of scale-resolving methods in industry has until recently been considered intractable.

Our vision with PyFR has been to develop and deliver an open-source Python framework based on the high-order accurate Flux Reconstruction (FR) approach \cite{Huynh2007} that enables real-world scale-resolving simulations to be undertaken in tractable time, at scale, by both academic and industrial practitioners---helping advance industrial CFD capabilities from their current ‘RANS plateau’. It is one of several international efforts towards industrial adoption of scale-resolving simulations, including Nektar++ \cite{cantwell2015nektar++,moxey2020nektar++}, hpMusic \cite{WANG2017579}, HiPSTAR \cite{sandberg2008development}, and charLES (now acquired and distributed for industrial usage by as Fidelity LES Solver by Cadence Inc) \cite{bres2018large,goc2021large}.

The first version of PyFR---v0.1.0---was released in late 2013 \cite{witherden2014pyfr}. It supported solving the compressible Euler and Navier--Stokes equations on unstructured grids of hexahedral elements, and was able to target both conventional CPUs and NVIDIA GPUs via a novel domain specific language based on Mako. In the past decade, there have been over 30 subsequent releases of PyFR, adding a wide range of new capabilities with a view to enabling industrial adoption,  culminating in the current release of PyFR v2.0.3 which is described in this paper.

%% file: fr.tex
\section{Flux Reconstruction}

%TODO add Figures.

\begin{figure*}[hbtp!]
  \centering
  \begin{subfigure}[b]{.98\linewidth}
    \centering
    {\includegraphics[width=0.8\textwidth]{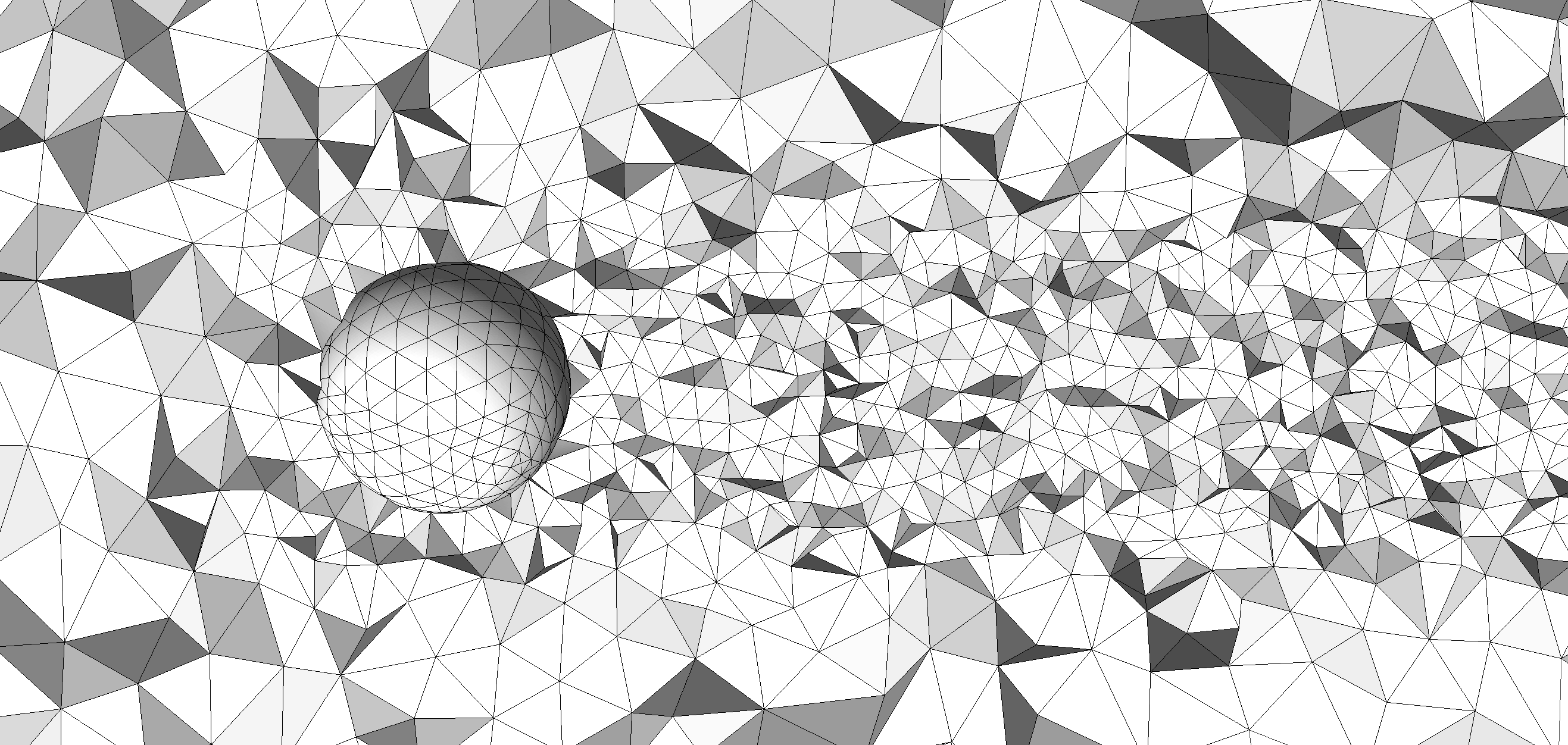}}%
    \caption{}
  \end{subfigure}
  \begin{subfigure}[b]{.48\linewidth}
    \centering
    \input{figs/tet-alpha-opt-surf-multi.tex}%
    \caption{}
  \end{subfigure}
  \begin{subfigure}[b]{.48\linewidth}
    \centering
    \input{figs/tet-alpha-opt.tex}%
    \caption{}
  \end{subfigure}
  \caption{Example of an unstructured curved-element tetrahedral mesh around a sphere (a), fourth-order $\alpha$-optimised flux points for a tetrahedron (b), where yellow and red indicate doubly and triply collocated points, respectively, and fourth-order $\alpha$-optimised solution points for a tetrahedron (c).}
  \label{fig:mesh}
\end{figure*}

The Flux Reconstruction (FR) approach \cite{Huynh2007} implemented in PyFR is a form of discontinuous spectral element method \cite{karniadakis2005spectral,hesthaven2007nodal}. As a brief overview, consider the first-order hyperbolic conservation law
\begin{equation}\label{eq:gov}
    \frac{\partial u_{\alpha}}{\partial t}
    +
    \nabla \cdot \mathbf{f}_{\alpha}
    =
    0,
\end{equation}
where $\alpha$ is the field variable index, $u_\alpha=u_\alpha(\mathbf{x},t)$ are the conservative field variables, and $\mathbf{f}_\alpha=\mathbf{f}_\alpha(u_\alpha)$ are the fluxes of $u_\alpha$. In order to solve Eq. \eqref{eq:gov} using FR in a domain $\mathbf{\Omega}$, one must tessellate the domain with $N$ non-overlapping conforming elements $\mathbf{\Omega}_{n}$ as 
\begin{equation}
    \mathbf{\Omega} = \bigcup_{n=1}^{N} \mathbf{\Omega}_{n}, \qquad
    \bigcap_{n=1}^{N}\mathbf{\Omega}_{n} = \varnothing,
\end{equation}
see, for example, Fig. \ref{fig:mesh}. Without loss of generality, we assume in this example all $\mathbf{\Omega}_{n}$ are of the same type.

Each element $\mathbf{\Omega}_n$ can then be mapped to a reference element $\hat{\mathbf{\Omega}}$ via a mapping function $\mathbf{\mathcal{M}}_\mathit{n}$ defined as
\begin{equation*}
    \mathbf{x} = \mathbf{\mathcal{M}}_\mathit{n}(\Tilde{\mathbf{x}}), \qquad
    \mathbf{\Tilde{x}} = \mathbf{\mathcal{M}}_\mathit{n}^{-1}(\mathbf{x}),
\end{equation*}
and geometric Jacobian matrices can be defined from the mapping functions as
\begin{alignat*}{4}
    \mathbf{J}_\mathit{n} &= J_\mathit{nij} &&= \frac{\partial \mathcal{M}_\mathit{ni}}{\partial \Tilde{x}_j}, \qquad
    && J_\mathit{n} &&= \det \mathbf{J}_\mathit{n}, \\
    \mathbf{J}^{-1}_\mathit{n} &= J^{-1}_\mathit{nij} &&= \frac{\partial \mathcal{M}^{-1}_\mathit{ni}}{\partial \Tilde{x}_j}, \qquad
    && J^{-1}_\mathit{n} &&= \det \mathbf{J}^{-1}_\mathit{n}=\frac{1}{J_\mathit{n}}.
\end{alignat*}
For each $\mathbf{\Omega}_{n}$, these Jacobian matrices can be used to transform Eq. \eqref{eq:gov} into reference element space as
\begin{equation}
    \frac{\partial u_{n\alpha}}{\partial t}
    +
    J^{-1}_\mathit{n}\Tilde{\nabla} \cdot \Tilde{\mathbf{f}}_{n\alpha}
    =
    0 \quad \text{and} \quad 
    \Tilde{\mathbf{f}}_\mathit{n\alpha} 
    = \Tilde{\mathbf{f}}_\mathit{n\alpha}(\Tilde{\mathbf{x}},t)
    = J_\mathit{n}(\Tilde{\mathbf{x}})\mathbf{J}^{-1}_\mathit{n}(\mathcal{M}_\mathit{n}(\Tilde{\mathbf{x}}))\mathbf{f}_\mathit{n\alpha}(\mathcal{M}_\mathit{n}(\Tilde{\mathbf{x}}),t),
\end{equation}
where $u_{n\alpha}$ and $\mathbf{f}_\mathit{n\alpha}$ are the solution and flux in $\mathbf{\Omega}_{n}$, respectively, and $\Tilde{\nabla} = \partial / \partial \Tilde{x}_i$.

We can proceed to define a set of solution points $\mathbf{\Tilde{x}}_{\zeta}^{(u)}$ in the reference element (see Fig. \ref{fig:mesh}), where $\zeta$ is the solution point index which satisfies $0 \leqslant \zeta < N^{(u)}$ and $N^{(u)}$ is the number of solution points in the reference element. Now a nodal basis set $\ell_{\zeta}^{(u)}(\mathbf{\Tilde{x}})$ can be defined in the reference element, where the nodal basis polynomials $\ell_{\zeta}^{(u)}$ satisfy $\ell_{\zeta}^{(u)}(\mathbf{\Tilde{x}}_{\sigma}^{(u)}) = \delta_{\zeta \sigma}$ where $\delta_{ij}$ is the Kronecker delta. We can also define a set of flux points $\mathbf{\Tilde{x}}_{\zeta}^{(f)}$ on the surface of the reference element (see Fig. \ref{fig:mesh}), where $\zeta$ is now the flux point index which satisfies $0 \leqslant \zeta < N^{(f)}$ and $N^{(f)}$ is the number of flux points on the surface of the reference element. These flux points are constrained such that flux points from adjoining elements always conform at element interfaces.

The first step in the FR approach is to obtain the discontinuous solution at each flux point $ u^{(f)}_\mathit{\sigma n\alpha}$ from the solution at the solution points $ u^{(u)}_\mathit{\zeta n\alpha}$ as
\begin{equation}\label{eq:disu}
    u^{(f)}_\mathit{\sigma n\alpha}
    =
    u^{(u)}_\mathit{\zeta n\alpha}\ell^{(u)}_\mathit{\zeta}
      (\Tilde{\mathbf{x}}^{(f)}_\mathit{\sigma}).
\end{equation}

The second step is to obtain a transformed common normal interface flux at each flux point $\tilde{f}^{C\mathit{(f_\perp)}}_\mathit{\sigma n \alpha}$ from the discontinuous solution at the flux point $u^\mathit{(f)}_\mathit{\sigma n}$, the discontinuous solution at the conforming flux point in the relevant adjoining element $u^{'\mathit{(f)}}_{\mathit{\sigma n}}$, and the surface normal at the flux point $\mathbf{n}^\mathit{(f)}_\mathit{\sigma n}$ as
\begin{equation}\label{eq:comm_flux}
    \tilde{f}^{C\mathit{(f_\perp)}}_\mathit{\sigma n \alpha}
    =
    \mathfrak{F}_\alpha
    (u^\mathit{(f)}_\mathit{\sigma n},
    u^{'\mathit{(f)}}_{\mathit{\sigma n}},
    \hat{\mathbf{n}}^\mathit{(f)}_\mathit{\sigma n}).
\end{equation}
where $\mathfrak{F}_\alpha$ is \textit{e.g.} an appropriate Riemann solver.

The third step is to calculate the transformed discontinuous flux at the solution points $\Tilde{\mathbf{f}}^\mathit{( u )}_\mathit{\sigma n \alpha}$ from the solution at solution points $\mathbf{u}^\mathit{( u )}_\mathit{\sigma n \alpha}$ using the system flux function. These values can then be used to calculate the transformed normal discontinuous flux at the flux points $\tilde{f}^{\mathit{(f_\perp)}}_\mathit{\sigma n \alpha}$ as
\begin{equation}\label{eq:disu}
    \tilde{f}^{\mathit{(f_\perp)}}_\mathit{\sigma n \alpha}
    =\mathbf{n}^\mathit{(f)}_\mathit{\sigma n}\cdot
    \Tilde{\mathbf{f}}^{(u)}_\mathit{\zeta n \alpha}\ell^{(u)}_\mathit{\zeta}
      (\Tilde{\mathbf{x}}^{(f)}_\mathit{\sigma}).
\end{equation}

Finally, the transformed divergence of the transformed continuous flux $(\Tilde{\nabla}\cdot\Tilde{\mathbf{f}})^\mathit{( u )}_\mathit{\zeta n \alpha}$ can be obtained by propagating the difference between $\tilde{f}^{C\mathit{(f_\perp)}}_\mathit{\sigma n \alpha}$ and $\tilde{f}^{\mathit{(f_\perp)}}_\mathit{\sigma n \alpha}$ at each flux point into the element using a flux correction function $\mathbf{g}^\mathit{( f )}_\mathit{\sigma}$, and combining with values of $\Tilde{\mathbf{f}}^\mathit{( u )}_\mathit{\sigma n \alpha}$ as
\begin{equation}\label{eq:FRfinal}
    (\Tilde{\nabla}\cdot\Tilde{\mathbf{f}})^\mathit{( u )}_\mathit{\zeta n \alpha}
    =
    \Big[
    \Tilde{\nabla}\cdot\Tilde{\mathbf{g}}^\mathit{( f )}_\mathit{\sigma}
        (\Tilde{\mathbf{x}})
        \big\{
        \mathfrak{F}_\alpha f^\mathit{(f_\perp)}_\mathit{\sigma n \alpha}
        - f^\mathit{(f_\perp)}_\mathit{\sigma n \alpha}
        \big\}
        + \Tilde{\mathbf{f}}^\mathit{( u )}_\mathit{\nu n \alpha}
            \cdot \Tilde{\nabla}\ell^\mathit{( u )}_\mathit{\nu}
                (\Tilde{\mathbf{x}})
    \Big]_{\Tilde{\mathbf{x}}=\Tilde{\mathbf{x}}^\mathit{( u )}_\mathit{\zeta}},
\end{equation}
which can then be used to update the solution at the solution points $ u^{(u)}_\mathit{\zeta n \alpha}$ via a suitable explicit time integration scheme.

%% file: figs/tet-alpha-opt-surf-multi.tex
\tdplotsetmaincoords{70}{110}
\begin{tikzpicture}[tdplot_main_coords,font=\scriptsize,scale=1.6,line join=round]
\coordinate (P1) at (-1,-1,-1);
\coordinate (P2) at (1,-1,-1);
\coordinate (P3) at (-1,-1,1);
\coordinate (P4) at (-1,1,-1);

\draw[thick] (P2) -- (P4) -- (P1);
\draw[thick] (P4) -- (P3);
\draw[thick] (P1) -- (P2) -- (P3) -- cycle;
\tdplottransformmainscreen{-1}{-1}{-1}
\shadedraw[tdplot_screen_coords, ball color=red] (\tdplotresx,\tdplotresy) circle (0.07);
\tdplottransformmainscreen{-0.654653670707977}{-1}{-1}
\shadedraw[tdplot_screen_coords, ball color=yellow] (\tdplotresx,\tdplotresy) circle (0.07);
\tdplottransformmainscreen{0}{-1}{-1}
\shadedraw[tdplot_screen_coords, ball color=yellow] (\tdplotresx,\tdplotresy) circle (0.07);
\tdplottransformmainscreen{0.654653670707977}{-1}{-1}
\shadedraw[tdplot_screen_coords, ball color=yellow] (\tdplotresx,\tdplotresy) circle (0.07);
\tdplottransformmainscreen{1}{-1}{-1}
\shadedraw[tdplot_screen_coords, ball color=red] (\tdplotresx,\tdplotresy) circle (0.07);
\tdplottransformmainscreen{-1}{-0.654653670707977}{-1}
\shadedraw[tdplot_screen_coords, ball color=yellow] (\tdplotresx,\tdplotresy) circle (0.07);
\tdplottransformmainscreen{-0.551583572090994}{-0.551583572090994}{-1}
\shadedraw[tdplot_screen_coords, ball color=green] (\tdplotresx,\tdplotresy) circle (0.07);
\tdplottransformmainscreen{0.103167144181987}{-0.551583572090994}{-1}
\shadedraw[tdplot_screen_coords, ball color=green] (\tdplotresx,\tdplotresy) circle (0.07);
\tdplottransformmainscreen{0.654653670707977}{-0.654653670707977}{-1}
\shadedraw[tdplot_screen_coords, ball color=yellow] (\tdplotresx,\tdplotresy) circle (0.07);
\tdplottransformmainscreen{-1}{0}{-1}
\shadedraw[tdplot_screen_coords, ball color=yellow] (\tdplotresx,\tdplotresy) circle (0.07);
\tdplottransformmainscreen{-0.551583572090994}{0.103167144181987}{-1}
\shadedraw[tdplot_screen_coords, ball color=green] (\tdplotresx,\tdplotresy) circle (0.07);
\tdplottransformmainscreen{0}{0}{-1}
\shadedraw[tdplot_screen_coords, ball color=yellow] (\tdplotresx,\tdplotresy) circle (0.07);
\tdplottransformmainscreen{-1}{0.654653670707977}{-1}
\shadedraw[tdplot_screen_coords, ball color=yellow] (\tdplotresx,\tdplotresy) circle (0.07);
\tdplottransformmainscreen{-0.654653670707977}{0.654653670707977}{-1}
\shadedraw[tdplot_screen_coords, ball color=yellow] (\tdplotresx,\tdplotresy) circle (0.07);
\tdplottransformmainscreen{-1}{1}{-1}
\shadedraw[tdplot_screen_coords, ball color=red] (\tdplotresx,\tdplotresy) circle (0.07);
\tdplottransformmainscreen{-1}{-1}{-0.654653670707977}
\shadedraw[tdplot_screen_coords, ball color=yellow] (\tdplotresx,\tdplotresy) circle (0.07);
\tdplottransformmainscreen{-0.551583572090994}{-1}{-0.551583572090994}
\shadedraw[tdplot_screen_coords, ball color=green] (\tdplotresx,\tdplotresy) circle (0.07);
\tdplottransformmainscreen{0.103167144181987}{-1}{-0.551583572090994}
\shadedraw[tdplot_screen_coords, ball color=green] (\tdplotresx,\tdplotresy) circle (0.07);
\tdplottransformmainscreen{0.654653670707977}{-1}{-0.654653670707977}
\shadedraw[tdplot_screen_coords, ball color=yellow] (\tdplotresx,\tdplotresy) circle (0.07);
\tdplottransformmainscreen{-1}{-0.551583572090994}{-0.551583572090994}
\shadedraw[tdplot_screen_coords, ball color=green] (\tdplotresx,\tdplotresy) circle (0.07);
\tdplottransformmainscreen{0.103167144181987}{-0.551583572090994}{-0.551583572090994}
\shadedraw[tdplot_screen_coords, ball color=green] (\tdplotresx,\tdplotresy) circle (0.07);
\tdplottransformmainscreen{-1}{0.103167144181987}{-0.551583572090994}
\shadedraw[tdplot_screen_coords, ball color=green] (\tdplotresx,\tdplotresy) circle (0.07);
\tdplottransformmainscreen{-0.551583572090994}{0.103167144181987}{-0.551583572090994}
\shadedraw[tdplot_screen_coords, ball color=green] (\tdplotresx,\tdplotresy) circle (0.07);
\tdplottransformmainscreen{-1}{0.654653670707977}{-0.654653670707977}
\shadedraw[tdplot_screen_coords, ball color=yellow] (\tdplotresx,\tdplotresy) circle (0.07);
\tdplottransformmainscreen{-1}{-1}{0}
\shadedraw[tdplot_screen_coords, ball color=yellow] (\tdplotresx,\tdplotresy) circle (0.07);
\tdplottransformmainscreen{-0.551583572090994}{-1}{0.103167144181987}
\shadedraw[tdplot_screen_coords, ball color=green] (\tdplotresx,\tdplotresy) circle (0.07);
\tdplottransformmainscreen{0}{-1}{0}
\shadedraw[tdplot_screen_coords, ball color=yellow] (\tdplotresx,\tdplotresy) circle (0.07);
\tdplottransformmainscreen{-1}{-0.551583572090994}{0.103167144181987}
\shadedraw[tdplot_screen_coords, ball color=green] (\tdplotresx,\tdplotresy) circle (0.07);
\tdplottransformmainscreen{-0.551583572090994}{-0.551583572090994}{0.103167144181987}
\shadedraw[tdplot_screen_coords, ball color=green] (\tdplotresx,\tdplotresy) circle (0.07);
\tdplottransformmainscreen{-1}{0}{0}
\shadedraw[tdplot_screen_coords, ball color=yellow] (\tdplotresx,\tdplotresy) circle (0.07);
\tdplottransformmainscreen{-1}{-1}{0.654653670707977}
\shadedraw[tdplot_screen_coords, ball color=yellow] (\tdplotresx,\tdplotresy) circle (0.07);
\tdplottransformmainscreen{-0.654653670707977}{-1}{0.654653670707977}
\shadedraw[tdplot_screen_coords, ball color=yellow] (\tdplotresx,\tdplotresy) circle (0.07);
\tdplottransformmainscreen{-1}{-0.654653670707977}{0.654653670707977}
\shadedraw[tdplot_screen_coords, ball color=yellow] (\tdplotresx,\tdplotresy) circle (0.07);
\tdplottransformmainscreen{-1}{-1}{1}
\shadedraw[tdplot_screen_coords, ball color=red] (\tdplotresx,\tdplotresy) circle (0.07);
\end{tikzpicture}

%% file: figs/tet-alpha-opt.tex
\tdplotsetmaincoords{70}{110}
\begin{tikzpicture}[tdplot_main_coords,font=\scriptsize,scale=1.6,line join=round]
\coordinate (P1) at (-1,-1,-1);
\coordinate (P2) at (1,-1,-1);
\coordinate (P3) at (-1,-1,1);
\coordinate (P4) at (-1,1,-1);

\draw[thick] (P2) -- (P4) -- (P1);
\draw[thick] (P4) -- (P3);
\draw[thick] (P1) -- (P2) -- (P3) -- cycle;
\tdplottransformmainscreen{-1}{-1}{-1}
\shadedraw[tdplot_screen_coords, ball color=blue] (\tdplotresx,\tdplotresy) circle (0.07);
\tdplottransformmainscreen{-0.654653670707977}{-1}{-1}
\shadedraw[tdplot_screen_coords, ball color=blue] (\tdplotresx,\tdplotresy) circle (0.07);
\tdplottransformmainscreen{0}{-1}{-1}
\shadedraw[tdplot_screen_coords, ball color=blue] (\tdplotresx,\tdplotresy) circle (0.07);
\tdplottransformmainscreen{0.654653670707977}{-1}{-1}
\shadedraw[tdplot_screen_coords, ball color=blue] (\tdplotresx,\tdplotresy) circle (0.07);
\tdplottransformmainscreen{1}{-1}{-1}
\shadedraw[tdplot_screen_coords, ball color=blue] (\tdplotresx,\tdplotresy) circle (0.07);
\tdplottransformmainscreen{-1}{-0.654653670707977}{-1}
\shadedraw[tdplot_screen_coords, ball color=blue] (\tdplotresx,\tdplotresy) circle (0.07);
\tdplottransformmainscreen{-0.551583572090994}{-0.551583572090994}{-1}
\shadedraw[tdplot_screen_coords, ball color=blue] (\tdplotresx,\tdplotresy) circle (0.07);
\tdplottransformmainscreen{0.103167144181987}{-0.551583572090994}{-1}
\shadedraw[tdplot_screen_coords, ball color=blue] (\tdplotresx,\tdplotresy) circle (0.07);
\tdplottransformmainscreen{0.654653670707977}{-0.654653670707977}{-1}
\shadedraw[tdplot_screen_coords, ball color=blue] (\tdplotresx,\tdplotresy) circle (0.07);
\tdplottransformmainscreen{-1}{0}{-1}
\shadedraw[tdplot_screen_coords, ball color=blue] (\tdplotresx,\tdplotresy) circle (0.07);
\tdplottransformmainscreen{-0.551583572090994}{0.103167144181987}{-1}
\shadedraw[tdplot_screen_coords, ball color=blue] (\tdplotresx,\tdplotresy) circle (0.07);
\tdplottransformmainscreen{0}{0}{-1}
\shadedraw[tdplot_screen_coords, ball color=blue] (\tdplotresx,\tdplotresy) circle (0.07);
\tdplottransformmainscreen{-1}{0.654653670707977}{-1}
\shadedraw[tdplot_screen_coords, ball color=blue] (\tdplotresx,\tdplotresy) circle (0.07);
\tdplottransformmainscreen{-0.654653670707977}{0.654653670707977}{-1}
\shadedraw[tdplot_screen_coords, ball color=blue] (\tdplotresx,\tdplotresy) circle (0.07);
\tdplottransformmainscreen{-1}{1}{-1}
\shadedraw[tdplot_screen_coords, ball color=blue] (\tdplotresx,\tdplotresy) circle (0.07);
\tdplottransformmainscreen{-1}{-1}{-0.654653670707977}
\shadedraw[tdplot_screen_coords, ball color=blue] (\tdplotresx,\tdplotresy) circle (0.07);
\tdplottransformmainscreen{-0.551583572090994}{-1}{-0.551583572090994}
\shadedraw[tdplot_screen_coords, ball color=blue] (\tdplotresx,\tdplotresy) circle (0.07);
\tdplottransformmainscreen{0.103167144181987}{-1}{-0.551583572090994}
\shadedraw[tdplot_screen_coords, ball color=blue] (\tdplotresx,\tdplotresy) circle (0.07);
\tdplottransformmainscreen{0.654653670707977}{-1}{-0.654653670707977}
\shadedraw[tdplot_screen_coords, ball color=blue] (\tdplotresx,\tdplotresy) circle (0.07);
\tdplottransformmainscreen{-1}{-0.551583572090994}{-0.551583572090994}
\shadedraw[tdplot_screen_coords, ball color=blue] (\tdplotresx,\tdplotresy) circle (0.07);
\tdplottransformmainscreen{-0.5}{-0.5}{-0.5}
\shadedraw[tdplot_screen_coords, ball color=blue] (\tdplotresx,\tdplotresy) circle (0.07);
\tdplottransformmainscreen{0.103167144181987}{-0.551583572090994}{-0.551583572090994}
\shadedraw[tdplot_screen_coords, ball color=blue] (\tdplotresx,\tdplotresy) circle (0.07);
\tdplottransformmainscreen{-1}{0.103167144181987}{-0.551583572090994}
\shadedraw[tdplot_screen_coords, ball color=blue] (\tdplotresx,\tdplotresy) circle (0.07);
\tdplottransformmainscreen{-0.551583572090994}{0.103167144181987}{-0.551583572090994}
\shadedraw[tdplot_screen_coords, ball color=blue] (\tdplotresx,\tdplotresy) circle (0.07);
\tdplottransformmainscreen{-1}{0.654653670707977}{-0.654653670707977}
\shadedraw[tdplot_screen_coords, ball color=blue] (\tdplotresx,\tdplotresy) circle (0.07);
\tdplottransformmainscreen{-1}{-1}{0}
\shadedraw[tdplot_screen_coords, ball color=blue] (\tdplotresx,\tdplotresy) circle (0.07);
\tdplottransformmainscreen{-0.551583572090994}{-1}{0.103167144181987}
\shadedraw[tdplot_screen_coords, ball color=blue] (\tdplotresx,\tdplotresy) circle (0.07);
\tdplottransformmainscreen{0}{-1}{0}
\shadedraw[tdplot_screen_coords, ball color=blue] (\tdplotresx,\tdplotresy) circle (0.07);
\tdplottransformmainscreen{-1}{-0.551583572090994}{0.103167144181987}
\shadedraw[tdplot_screen_coords, ball color=blue] (\tdplotresx,\tdplotresy) circle (0.07);
\tdplottransformmainscreen{-0.551583572090994}{-0.551583572090994}{0.103167144181987}
\shadedraw[tdplot_screen_coords, ball color=blue] (\tdplotresx,\tdplotresy) circle (0.07);
\tdplottransformmainscreen{-1}{0}{0}
\shadedraw[tdplot_screen_coords, ball color=blue] (\tdplotresx,\tdplotresy) circle (0.07);
\tdplottransformmainscreen{-1}{-1}{0.654653670707977}
\shadedraw[tdplot_screen_coords, ball color=blue] (\tdplotresx,\tdplotresy) circle (0.07);
\tdplottransformmainscreen{-0.654653670707977}{-1}{0.654653670707977}
\shadedraw[tdplot_screen_coords, ball color=blue] (\tdplotresx,\tdplotresy) circle (0.07);
\tdplottransformmainscreen{-1}{-0.654653670707977}{0.654653670707977}
\shadedraw[tdplot_screen_coords, ball color=blue] (\tdplotresx,\tdplotresy) circle (0.07);
\tdplottransformmainscreen{-1}{-1}{1}
\shadedraw[tdplot_screen_coords, ball color=blue] (\tdplotresx,\tdplotresy) circle (0.07);
\end{tikzpicture}

%% file: additions.tex
\section{New Capabilities}

%\begin{table}[tbhp]
%    \centering
%    \begin{tabular}{ccc}
%        \toprule
%         & 0.1.0 & 2.0.3 \\ \midrule
%        Governing equations & Euler, Navier--Stokes & \makecell{Euler, Navier--Stokes, \\ incompressible Euler, incompressible Navier-Stokes}\\
%        Element types & hex, tri, quad & tets, hex, tri, quad, pri, pyr \\
%        Time stepping & Euler, RK4, DOPRI5 & \makecell{Euler, RK34, RK4, RK45, TVD-RK3,\\ backward-Euler, SDIRK33, SRIRK43} \\
%        Backends & CUDA, OpenMP &  CUDA, HIP, Metal, OpenCL, OpenMP \\ 
%        Plugins & 0 & 13\\
%        \bottomrule
%    \end{tabular}
%    \caption{Overview}
%\end{table}

\subsection{Cross-Platform Performance}

\begin{figure}
    \centering
    \includegraphics[width=\textwidth]{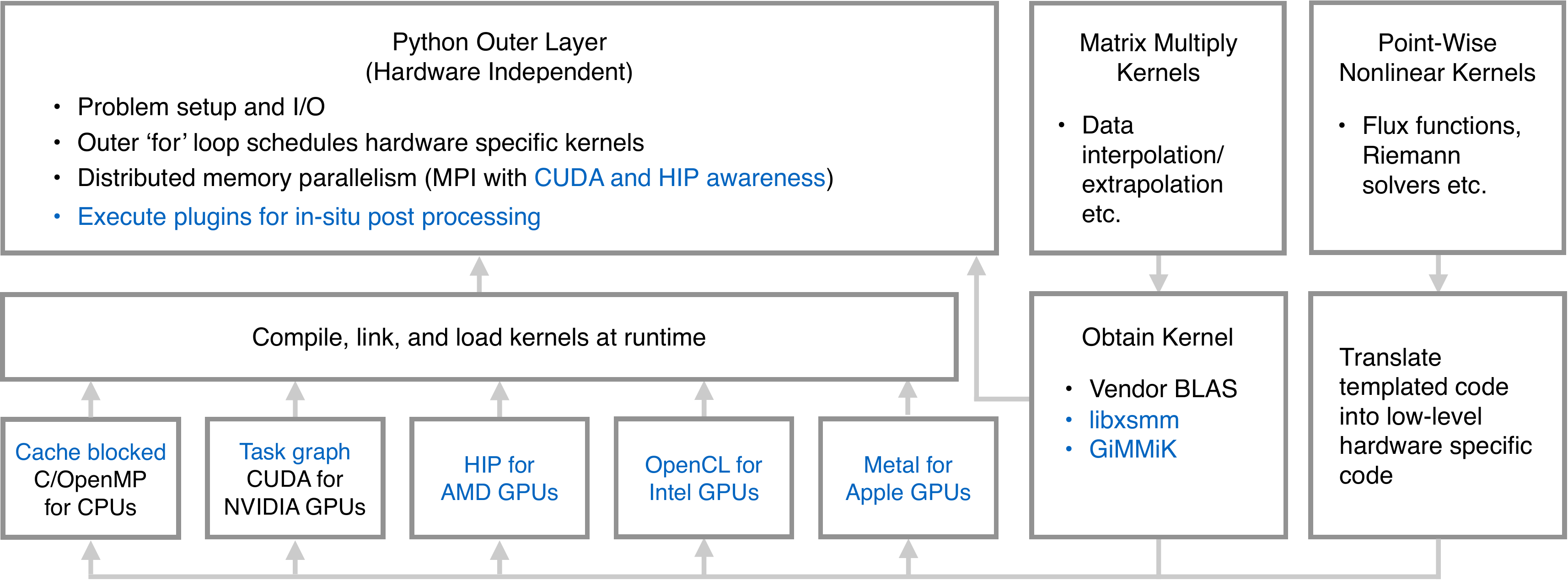}
    \caption{Overview of how PyFR achieves cross-platform performance.  New functionality is marked in blue.}
    \label{fig:pyfr-structure}
\end{figure}

The cross-platform performance of PyFR is enabled through \emph{backends} which can utilise various matrix multiplication kernels and a Mako derived domain specific language (DSL) which achieves complete feature parity across all backends, as per \cref{fig:pyfr-structure}.

\paragraph{Backends.}
  PyFR v0.1.0 had a \texttt{C/OpenMP} backend for CPUs and a \texttt{CUDA} backend for NVIDIA GPUs.  However, since 2013 additional vendors have entered the high-end GPU market including AMD, Intel, and even Apple.  As such, PyFR v2.0.3 now contain an additional \texttt{HIP} backend for AMD GPUs, an \texttt{OpenCL} backend for all GPUs, and a \texttt{Metal} backend for Apple GPUs.

\paragraph{DSL.}
The capabilities and performance of the DSL have been improved in PyFR v2.0.3.  In particular, there is language-level support for \emph{reductions}.  A kernel argument can now be annotated according to \texttt{reduce(op)} where \texttt{op} is a reduction operator such as \texttt{min}.  Whatever value the kernel assigns to this argument will then be automatically and safely reduced with its current value in memory.  On the performance side, the DSL is now capable of detecting situations where read-only kernel arguments are likely to be subject to reuse.  When running on GPU platforms, the DSL will automatically take care of loading these arguments into shared memory.  This helps to reduce pressure on the L1 and L2 caches. 

\paragraph{Matrix multiplications.}
Many operations within an FR time-step can be cast in the form of
\[
 \boldsymbol{\mathsf{C}} \gets \boldsymbol{\mathsf{A}}\boldsymbol{\mathsf{B}} + \beta \boldsymbol{\mathsf{C}},
\]
where $\boldsymbol{\mathsf{A}}$ is a constant operator matrix, $\boldsymbol{\mathsf{B}}$ is an input state matrix, and $\boldsymbol{\mathsf{C}}$ is an output state matrix.  PyFR v0.1.0 simply offloaded these operations to a platform-specific dense BLAS library such as cuBLAS or OpenBLAS.  However, when operating on elements with a tensor-product structure, the operator matrices can exhibit a significant degree of sparsity.  This can lead to suboptimal performance in cases where the arithmetic intensity of the operation is beyond that of the underlying hardware.

This issue has been addressed by incorporating additional matrix-multiplication providers into PyFR v2.0.3.  When running on with the CUDA and HIP backends, PyFR v2.0.3 will use the GiMMiK \cite{wozniak2016gimmik} library to generate a suite of bespoke fully-unrolled kernels for each $\boldsymbol{\mathsf{A}}$.  During the code generation process, GiMMiK automatically elides multiplications through by zero, thus reducing the arithmetic intensity of the operation.  The generated kernels are then competitively benchmarked against those provided by the dense BLAS library, with PyFR automatically selecting the fastest kernel for each operation.  This auto-tuning is performed autonomously by PyFR at run-time and does not require any direction from the user.  When running on CPUs, a similar result is accomplished through the use of libxsmm \cite{heinecke2016libxsmm} which includes its own built-in support for automatically choosing between dense and sparse kernels.

Finally, in situations where the flux points are a strict subset of the solution points, additional logic has been incorporated into PyFR to avoid the need for multiplications entirely.  This leads to further memory and memory bandwidth savings.

\paragraph{Data layout.}

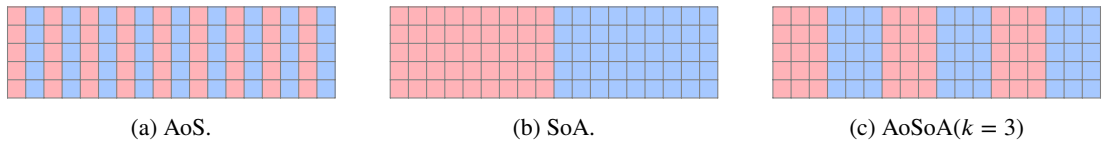
\begin{figure}
 \centering
 \begin{subfigure}[b]{0.3\textwidth}%
  \centering%
  \input{figs/fr-mem-aos.tex}%
  \caption{AoS.}
 \end{subfigure}
 \begin{subfigure}[b]{0.3\textwidth}%
  \centering%
  \input{figs/fr-mem-soa.tex}%
  \caption{SoA.}
 \end{subfigure}
 \begin{subfigure}[b]{0.3\textwidth}%
   \centering%
   \input{figs/fr-mem-aosoa.tex}%
   \caption{AoSoA($k = 3$)}
 \end{subfigure}
 \caption{\label{fig:fr-mem}Data layout methodologies for packing multiple field variables into the rows of a matrix.}
\end{figure}

The primary data structure in PyFR is an $m$ by $n$ row-major matrix where, up to padding, $m$ is proportional to the number of solution/flux points and $n$ is equal to the product of the number of elements and the number of field variables.  It follows that there is a degree of freedom regarding how these field variables are packed along a row.  This can be characterised by the stride $\Delta j$ between two subsequent field variables.  The choice of $\Delta j = 1$ results in an array of structures (AoS) arrangement, $\Delta j = N_E$ where $N_E$ is the number of elements results in a structure of arrays (SoA) arrangement, and $\Delta j = k$ results in a hybrid array of structure of arrays (AoSoA) approach.  An illustration of these arrangements can be seen in \cref{fig:fr-mem}.

For simplicity, PyFR v0.1.0 used an SoA approach.  However, although this structure is readily amenable to vectorisation, it has some limitations.  Firstly, the large stride between field variables decreases the efficiency of caches since adjacent field variables are unlikely to reside in the same cache line.  Secondly, it is not friendly to hardware pre-fetchers: an SoA structure with $v$ field variables appears to a CPUs pre-fetcher like $v$ \emph{separate} arrays.  Since CPUs are only capable of pre-fetching a finite number of data streams, this can lead to stalls.  Finally, given a pointer to one field variable at one point, it is not possible to access the next field variable unless one also knows $N_E$.  To avoid these issues, PyFR v2.0.3 employs the more sophisticated AoSoA packing.  The value of $k$ is chosen automatically by the backend based on the vector length of the underlying hardware.

Additionally, when running on CPUs, PyFR v2.0.3 incorporates an additional level of blocking.  Rather than allocating a single row-major matrix with $n$ columns, the C/OpenMP backend instead allocates $q$ smaller matrices each with ${\sim} n/q$ columns.  The value of $q$ is chosen to ensure that an entire block can easily remain resident in local caches and serves to further improve data locality.

\paragraph{Task graphs.}
Strong scaling has also been improved by adding first-class support for \emph{task graphs}.  The idea is to exploit the fact that PyFR, as with many scientific codes, repeatedly calls the same sequence of kernels.  By treating each kernel as a vertex in a graph and the dependencies between kernels as edges, it is possible---on an \emph{a priori} basis---to form a task graph corresponding to a single right-hand side evaluation.  This has two key advantages:
\begin{enumerate}
    \item It presents the underlying runtime with extra opportunities for extracting parallelism by enabling it to safely identify kernels which can be run in parallel.
    \item It enables a substantial reduction in interface overhead since, once constructed, task graphs can be launched with just a single function call as opposed to one function call per kernel.
\end{enumerate}

\begin{figure}
    \centering
    \includegraphics[height=17.5cm]{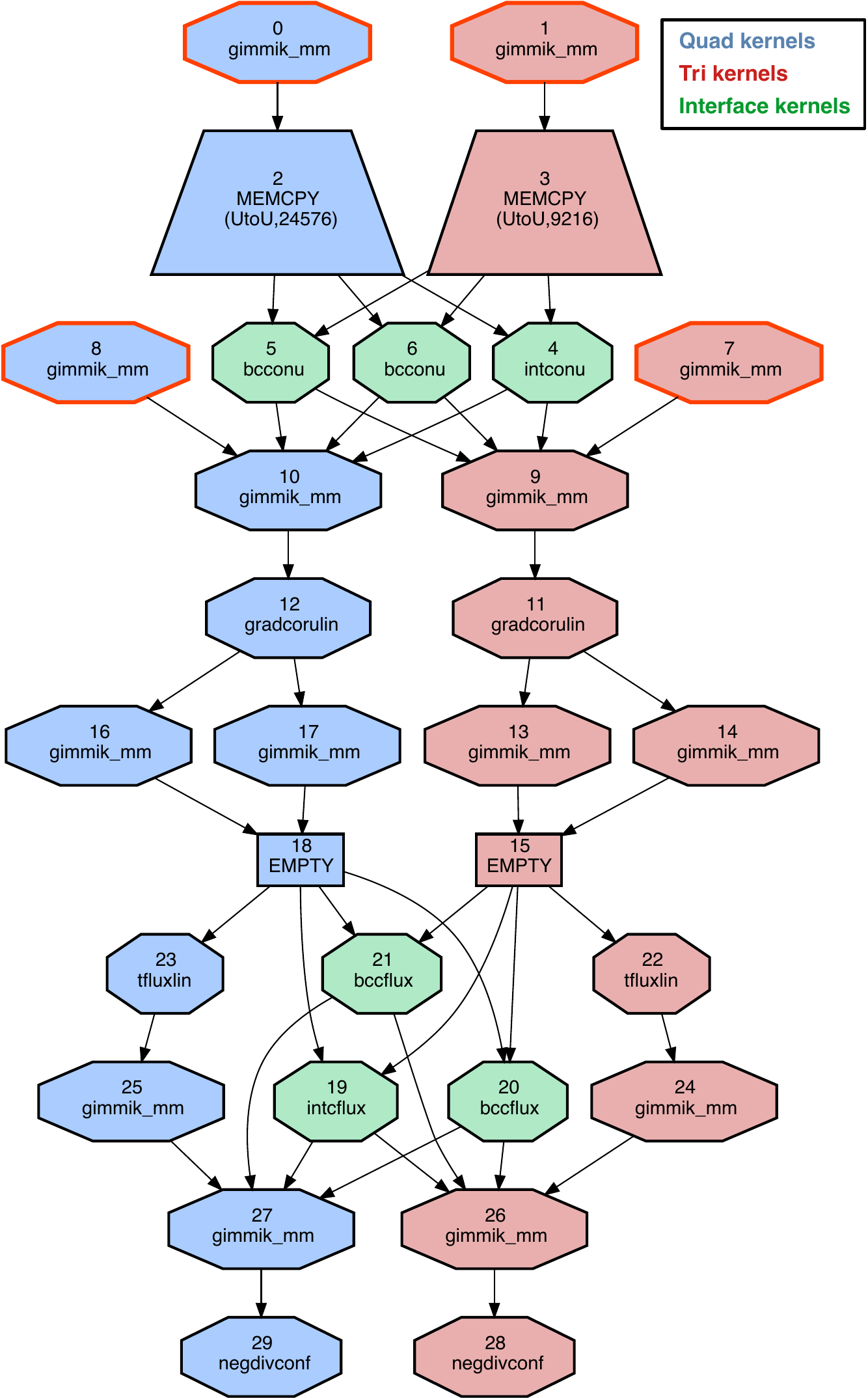}
    \caption{A task graph generated by the NVIDIA \texttt{cuGraphDebugDotPrint} API for a 2D mixed-element Navier--Stokes test case.  Blue shading indicates kernels for quadrilateral elements, red shading kernels for triangular elements, and green shading for interface kernels. The four root nodes of the graph are marked with red borders.}
    \label{fig:graph}
\end{figure}

An example of a task graph for a 2D Navier--Stokes simulation on a mixed-element grid can be seen in \cref{fig:graph}.  Looking at the graph, we observe that there are four root nodes.  As these root nodes are---by definition---independent, it is possible for all four of the kernels to be executed in parallel.  Similarly, we observe that the three \texttt{bcconu} boundary condition kernels are also independent, such that these can also be executed in parallel.  Indeed, careful inspection of the graph shows that there are \emph{always} at least two kernels which may be executed at the same time.  When exploited by a backend, this parallelism can improve GPU utilisation which, in turn, leads to improved strong scaling.

On the CUDA backend, PyFR task graphs map directly onto native CUDA graphs.  Since the NVIDIA A100 generation, there is hardware support for task graph acceleration, enabling further reductions in overhead.  While HIP does support task graphs, preliminary studies show their performance to be inferior to launching kernels directly.  As such, task graphs on the HIP backend are emulated by submitting kernels to a stream in a serial fashion.  A similar approach is used on Metal.  On OpenCL, task graphs are emulated using out-of-order queues and events.  This enables the runtime to identify and exploit inter-kernel parallelism but does not decrease API overhead.

To demonstrate the benefits of native task graphs, we consider the 2D Incompressible Cylinder Flow test case \href{https://github.com/PyFR/PyFR-Test-Cases/tree/main/2d-inc-cylinder}{\texttt{2d-inc-cylinder}} available from the PyFR Test Case repository on GitHub.  This is small mixed-element case has a high API overhead, and without task graphs, the run-time on an NVIDIA V100 GPUs is \qty[exponent-mode = fixed, fixed-exponent = 0]{256}{s}.  However, with task graphs, this reduces to \qty[exponent-mode = fixed, fixed-exponent = 0]{122}{s}.

%TODO add cache blocking references

\paragraph{Cache blocking.}
A powerful means of reducing the memory bandwidth requirements of a code on conventional CPUs is \emph{cache blocking} \cite{akkurt2022cache}.  The idea is to improve data locality by changing the order in which kernels are called.  An example of this can be seen in \cref{fig:cblock} which shows how a pair of array addition kernels can be rearranged to reduce bandwidth requirements.  A key advantage of cache blocking compared with alternative approaches, such as kernel fusion, is that the kernels themselves do not require modification; all that changes is the arguments to the kernels.

\begin{figure}
    \hfil
    \begin{subfigure}[b]{0.4\textwidth}
     \centering
\begin{lstlisting}[language=C]
// Kernel 1
for (int i = 0; i < n; i++)
    a[i] += b[i];

// Kernel 2
for (int i = 0; i < n; i++)
    a[i] += c[i];
€
\end{lstlisting}
     \caption{Without blocking; bandwidth ${\sim}6n$.}
    \end{subfigure}\hfil
    \begin{subfigure}[b]{0.5\textwidth}
     \centering
\begin{lstlisting}[language=C]
// Blocking loop with b = block size
for (int j = 0; j < n; j += b) {
    // Kernel 1
    for (int i = j; i < j + b; i++)
        a[i] += b[i];

    // Kernel 2
    for (int i = j; i < j + b; i++)
        a[i] += c[i];
}
\end{lstlisting}
     \caption{With blocking; bandwidth ${\sim}4n$.}
    \end{subfigure}\hfil
    \caption{Example of how cache blocking can be applied to a pair of array addition kernels.  In the blocked version when the second kernel updates \texttt{a[i]} it will hit in cache, thus saving a write to and read back from main memory.}
    \label{fig:cblock}
\end{figure}

Historically, cache blocking has not been viable for high-order codes due to the size of the intermediate arrays which are generated by kernels.  For example, an Intel Ivy Bridge CPU core from 2013 only has \qty[exponent-mode = fixed, fixed-exponent = 0]{256}{\kibi\byte} of L2 cache which is shared between executable code and data.   As a point of reference, for the Euler equations, storing the solution and flux for just eight $\wp = 4$ hexahedra at double precision requires \qty[exponent-mode = fixed, fixed-exponent = 0]{160}{\kilo\byte}.  Since 2016, however, there has been a marked increase in the size of private caches, with Intel Golden Cove CPU cores having \qty{2}{\mebi\byte}.  The specifics involved in cache blocking FR are detailed in \cite{akkurt2022cache, 10.52843/gpnpwx} and can improve performance by a factor of two.  Within PyFR, cache blocking is accomplished by calling auxiliary methods on task graphs stating which kernels in the graph are suitable for blocking transformations.  The interface also contains support for eliminating temporary arrays which can further improve performance.

\paragraph{Multi-node capabilities.}
Distributed memory parallelism is accomplished via MPI using the mpi4py wrappers \cite{dalcin2005mpi,dalcin2008mpi}.  As the message format is standardised across all backends, it is possible for different ranks to employ different backends, thus enabling heterogeneous computing from a homogeneous codebase \cite{witherden2015heterogeneous}.  In order to improve scalability, the backend interface in PyFR v2.0.3 has been enhanced to allow backends to directly pass GPU device pointers to MPI routines.  As such, PyFR v2.0.3 is fully capable of exploiting GPUDirect RDMA on NVIDIA platforms via CUDA Aware MPI, along with its analogue on AMD platforms via HIP Aware MPI.  The impact of this technology depends on both the underlying hardware and the degree to which a simulation is strong scaled.  In the most extreme cases, twofold performance improvements have been observed when running on clusters of NVIDIA A100 GPUs \cite{mishra2023scaling}.

\subsection{Numerical Stability}

PyFR is often used to conduct under-resolved DNS (uDNS), also referred to as ILES of turbulent flow.  On account of this under-resolution, the FR scheme is subject to aliasing-driven instabilities, which can cause the simulation to diverge \cite{jameson2012non}.  Additionally, FR schemes exhibit instabilities when solutions contain discontinuities such as shocks.  PyFR v0.1.0 had no specialised capabilities for handling either scenario, beyond simply increasing grid resolution. However, as of PyFR v2.0.3, there are now four separate stabilisation strategies available.

\paragraph{Modal filtering.}
The simplest stabilisation technique in PyFR v2.0.3 is modal filtering, wherein high-order modes of the solution are periodically filtered as outlined in \cite{hesthaven2007nodal}.  This approach is conservative and numerically inexpensive.  However, it is an indiscriminate approach - with the filtering applied uniformly across the domain irrespective of whether it is required, and it exposes several free parameters including the filter strength, filter frequency and cut-off modes. 

\paragraph{Anti-aliasing.}
As noted in \cite{jameson2012non}, the origin of aliasing-driven instabilities is the use of a collocation-type projection of the fluxes.  PyFR v2.0.3 resolves this issue by using quadrature to perform a least-squares projection of the flux instead.  To do this, PyFR employs a series of state-of-the art quadrature rules generated using Polyquad \cite{witherden2015identification}, which out-perform those in literature.  Although computationally expensive relative to simply performing a collocation projection of the fluxes, studies have shown the results to be markedly superior to those produced by modal filtering \cite{park2017high}. 

\paragraph{Artificial viscosity.}
Primarily intended for shock capturing, artificial viscosity is another stabilisation approach provided by PyFR v2.0.3, which dynamically adds extra viscosity into elements whose solutions are exhibiting Gibbs-type phenomena.  Based around the widely adopted approach of \cite{persson2006sub}, the method is functional but, as with modal filtering, requires a degree of parameterisation.  Additionally, whilst the additional kernels are not particularly expensive---at least within the context of an advection-diffusion type problem such as the Navier--Stokes equations---the process of adding viscosity can have a negative impact on the maximum stable explicit time step. As such,  the overall cost of the approach can be high.

\paragraph{Entropy filtering.}
The final approach provided by PyFR v2.0.3 for stabilisation and shock capturing is entropy filtering \cite{dzanic2022positivity,dzanic2023positivity,10.52843/cassyni.pvy6c0,Trojak2024}.  This is based around selectively applying a modal filter to elements which violate positivity of density, positivity of pressure or a minimum entropy condition.  The filter strength is determined iteratively on a per-element basis, with the goal being to apply as little filtering as possible.  As the indicators for instability are physics-based, this method does not typically require any explicit parameterisation. The utility of the approach is demonstrated in the 2D Double Mach Reflection test case \href{https://github.com/PyFR/PyFR-Test-Cases/tree/main/2d-double-mach-reflection}{\texttt{2d-double-mach-reflection}} and the 2D Viscous Shock Tube test case \href{https://github.com/PyFR/PyFR-Test-Cases/tree/main/2d-viscous-shock-tube}{\texttt{2d-viscous-shock-tube}} available from the PyFR Test Case repository on Github.

\subsection{Mixed Elements and Domain Decomposition}

PyFR v0.1.0 only included support for three element types: quadrilaterals and triangles in two dimensions and hexahedra in three dimensions.  Given the difficulties of all-hexahedral meshing around complex geometries, this represented a significant limitation.  PyFR v2.0.3 addresses this limitation by adding in complete support for prisms and tetrahedra and partial support for pyramids.  Specifically, the pyramid support requires that the quadrilateral base be affine.  Given that a major application for pyramids is as a transition layer between a tetrahedral near-field and a hexahedral far-field, this restriction is relatively minor.

One practical complication which arises when running on mixed grids is domain decomposition.  The relative performance of different element types is affected by around half a dozen simulation parameters including: the polynomial order, location of solution and flux points and use of anti-aliasing, to name but three.  A consequence of this is that when partitioning a grid, it is not possible to employ a single set of element weighting factors.  Employing incorrect weighting factors can lead to load imbalances which negatively impact strong scaling.  Compared with v0.1.0, PyFR v2.0.3 contains two major improvements in this area.

Firstly, whereas v0.1.0 required grids to be partitioned by the mesh generation software, v2.0.3 includes built-in support for partitioning and re-partitioning both mesh and solution files.  This is accomplished by having PyFR call out to the METIS \cite{karypis1997metis} and SCOTCH \cite{chevalier2008pt} libraries.  Using this functionality, it is relatively simple to experiment with different weightings and change them in concert with the simulation parameters; for example, when restarting a simulation at a higher polynomial order, this functionality can be used to appropriately re-weight the mesh.  Moreover, to aid this process, PyFR also includes support for tracking MPI wait times.  This information can be used to identify load imbalances between domains, which the user can then employ to derive more appropriate weights.

\begin{figure}
    \centering
    \begin{subfigure}{0.45\textwidth}
    \centering
        \includegraphics[width=\textwidth]{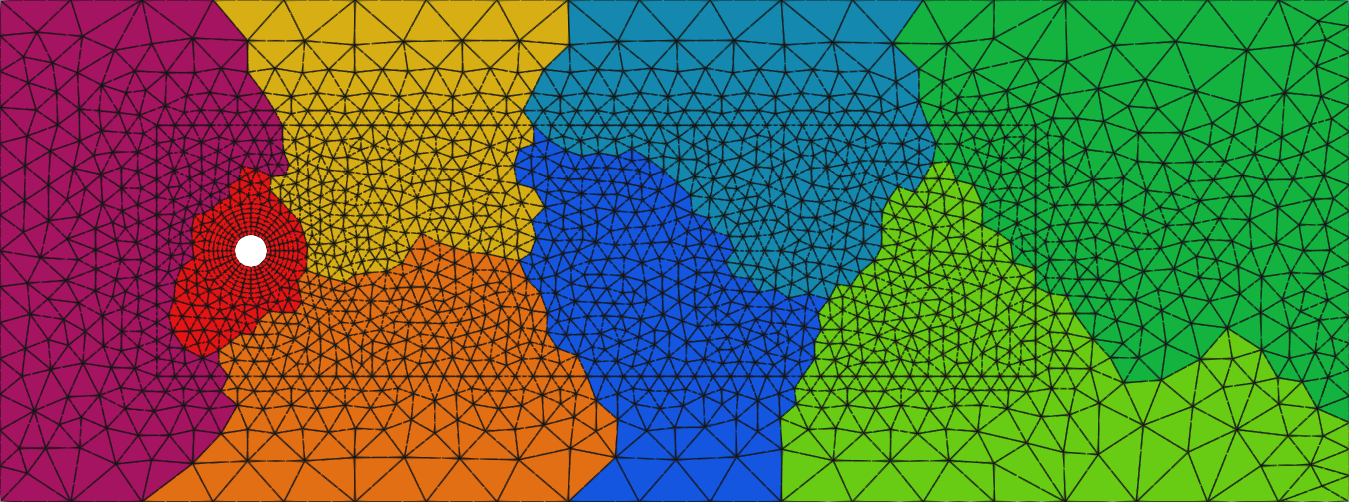}
        \caption{Weighted.}
    \end{subfigure}
    \begin{subfigure}{0.45\textwidth}
    \centering
        \includegraphics[width=\textwidth]{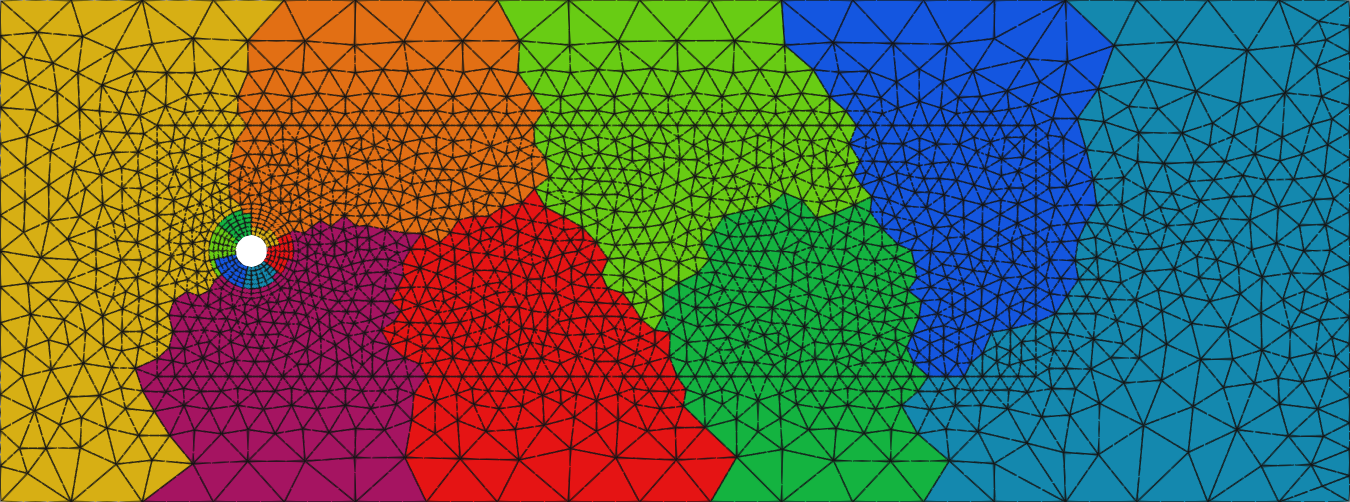}
        \caption{Balanced.}
    \end{subfigure}
    \caption{Partitionings of a mixed grid with a quadrilateral boundary layer and triangular far-field partitioned into eight parts.  For the weighted strategy, each quadrilateral was assigned $3/2$ the weight of a triangle.}
    \label{fig:partitioning}
\end{figure}

Secondly, there is also support for \emph{balanced} partitioning wherein PyFR attempts to assign the same number of elements of each type to each domain.  This ensures optimal load balancing irrespective of the relative performance differential between element types.  However, as element types are not uniformly distributed throughout the domain---for example one might have a prismatic boundary layer and a tetrahedral wake---balanced partitioning can lead to partitions becoming non-contiguous when the number of partitions is large.  Examples of weighted and balanced partitioning can be seen in \cref{fig:partitioning} for the 2D Incompressible Cylinder Flow test case \href{https://github.com/PyFR/PyFR-Test-Cases/tree/main/2d-inc-cylinder}{\texttt{2d-inc-cylinder}} available from the PyFR Test Case repository on GitHub.

\subsection{Curved Elements}

To realise the benefits of high-order schemes, it is necessary to employ grids which, by finite volume standards, are relatively coarse.  In order to still accurately represent the underlying geometry, it is therefore important for the elements themselves to be curved.  This is accomplished by associating metric terms---which take the form of a $2 \times 2$ or $3 \times 3$ matrix---within each element.

PyFR v0.1.0 employed the so-called cross-product metric.  However, with this approach, the polynomial order of the spatial metric is twice that of the shape function for curved grids---possibly exceeding the order of the solution basis. If this is the case, then the metric terms may not be discretised accurately due to truncation and aliasing errors. In particular, the divergence of the approximated metric terms may become non-zero, which results in a lack of free-stream preservation, \textit{i.e.} the solver cannot maintain a uniform free-stream flow solution.

\begin{figure}
    \centering
    \begin{subfigure}{0.45\textwidth}
    \centering
        \includegraphics[width=\textwidth]{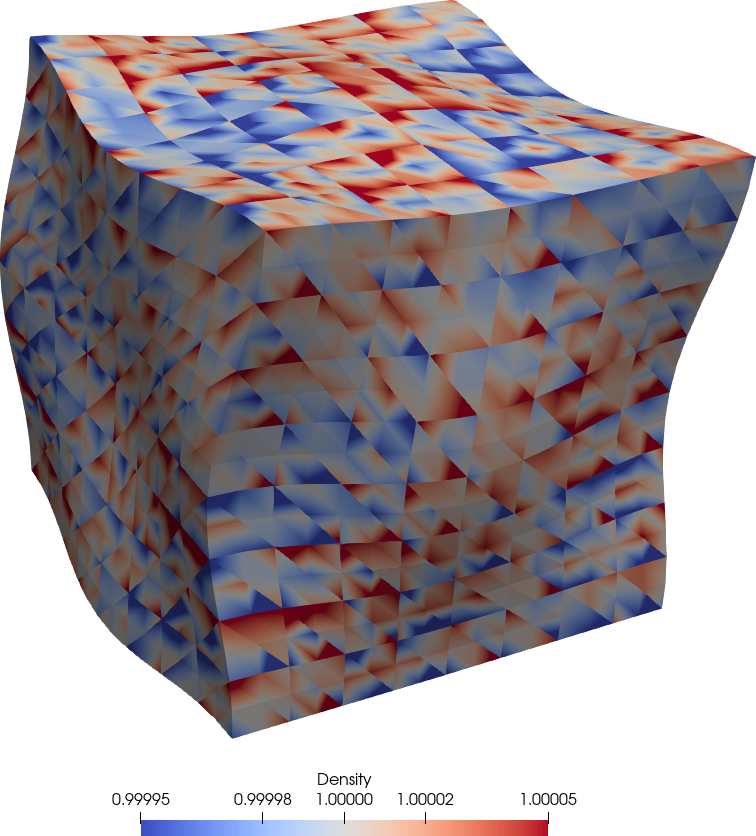}
        \caption{Cross-product metric as used in PyFR v0.1.0.}
    \end{subfigure}\hfill
    \begin{subfigure}{0.45\textwidth}
    \centering
        \includegraphics[width=\textwidth]{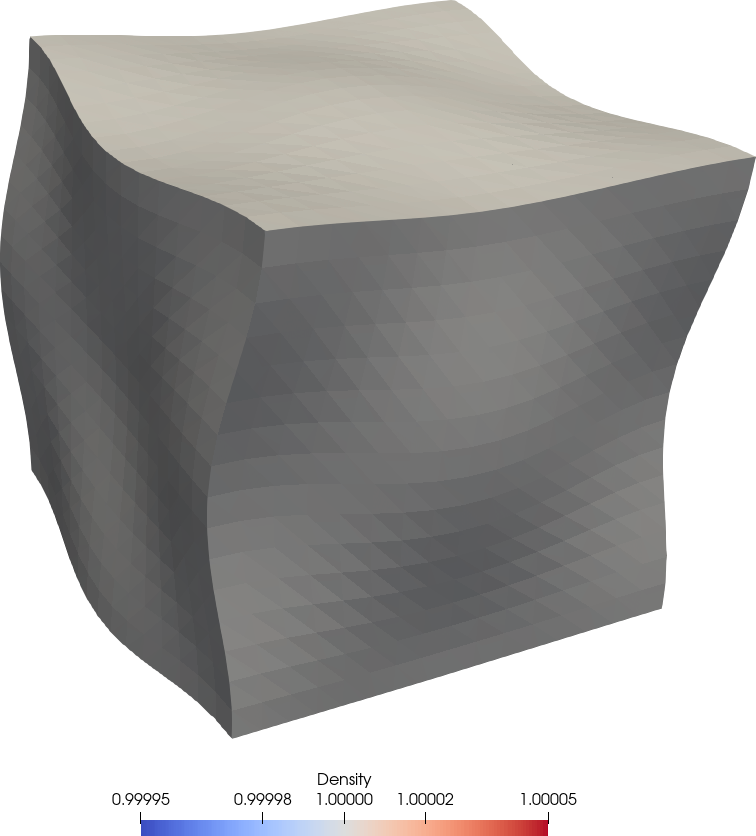}
        \caption{Conservative metric as used in PyFR v2.0.3.}
    \end{subfigure}
    \caption{Simulation of free-stream flow using second-order solution polynomials on a cubically-curved tetrahedral mesh with cross-product metric as used in PyFR v0.1.0 (a), and conservative metric as used in PyFR v2.0.3 (b).}
    \label{fig:fsp}
\end{figure}

To overcome this issue, PyFR v2.0.3 instead employs the conservative metric~\cite{kopriva2006metric,abe2015freestream}, which preserves a uniform free-stream flow even when discretised.  Specifically, this approach constructs the metric terms as the curl of the function, thus ensuring they are always divergence-free irrespective of any errors in the function approximation.  The approach greatly increases the robustness of PyFR when running on curved grids.  An example of its impact can be seen in \cref{fig:fsp}.

Furthermore, in many real-world grids, only elements in and around the boundary layer are actually curved.  PyFR v2.0.3 takes advantage of this fact by identifying linear elements and, in lieu of computing metric terms for each solution point on an \emph{a priori} basis, instead determines them on the fly based off the geometry of the element.  This can lead to a substantial saving in memory bandwidth.  For example, in a $\wp = 4$ hexahedral element there are $(\wp + 1)^3 = 125$ solution points and hence $3^2\times 125 = 1,125$ metric terms.  However, if the element is linear, then the metric terms are entirely determined by the corner vertices which only involve $3 \times 8 = 24$ terms.

\subsection{Adaptive Time Stepping}

When using explicit time stepping, the run-time of a simulation is directly proportional to the time step size.  However, for non-linear problems, it can be challenging to accurately estimate the maximum stable step size.  PyFR v2.0.3 avoids these issues by including support for low storage Runge--Kutta methods with \emph{embedded pairs}.  These make it possible to inexpensively obtain an estimate for the numerical error incurred when taking a time step \cite{heirer1993solving}.  Once suitably normalised, this error is then used to decide if a time step should be accepted or rejected.  Moreover, it is also be used to adapt the step size; increasing for accepted steps and decreasing for rejected steps.

\subsection{Incompressible Euler and Navier--Stokes}

PyFR v0.1.0 included a compressible Euler and Navier--Stokes solver. In PyFR v2.0.3, support has also been added for the incompressible Navier--Stokes equations.  This is accomplished via a combination of the artificial compressibility method of \cite{chorin1997numerical} with the dual-time approach of \cite{jameson1991time}.  The result is an iterative scheme which builds extensively upon the fast residual evaluation capability of PyFR.  Convergence is accelerated through a combination of polynomial multigrid \cite{loppi2018high} and variable local time stepping \cite{loppi2019locally}.

\subsection{File Formats}

The mesh and solution file formats for PyFR v0.1.0 were based around NumPy \texttt{.npy} and \texttt{.npz} files.  Although simple to read and write from Python, they are difficult to access from other environments.  Moreover, the formats themselves had no provisions for parallel I/O.  For these reasons, PyFR v2.0.3 employs a new set of file formats based around the industry-standard HDF5 \cite{folk2011overview,collette2013python}.  A key advantage of the HDF5 format is its hierarchical nature and the ability to attach arbitrary attributes to most data sets.  Moreover,  data arrays stored by PyFR use 64-bit rather than 32-bit integers.  This enables a partition to have in excess of four billion elements and serves to further future-proof the format for at least the next decade.

To aid in reproducibility, PyFR v2.0.3 solution files embed \emph{all} of the configuration files that have been employed in the simulation up until the current time.  This makes it possible to account for the common situation wherein a simulation is started with one configuration file, run for a period of time, and then restarted with a different configuration.

Output format support has also been enhanced since PyFR v0.1.0.  Specifically, PyFR v2.0.3 now supports exporting high-order VTU files.  This enables the high-order nature of the solution to be preserved throughout more of the post-processing pipeline.  Furthermore, there is also support for generating parallel VTU files which can be more efficient in multiprocessing environments.

\subsection{Plugin Architecture}

The PyFR \emph{plugin} infrastructure provides a lightweight means of adding new capabilities to the code base.  Written in pure Python, plugins are capable of adding new command line arguments, periodically post-processing the solution, and adding source-terms to the solver. Examples of a selection of plugins provided with PyFR v2.0.3 are detailed below.

\paragraph{Point sampling.}
The \texttt{soln-plugin-sampler} is capable of periodically sampling a set of points in the domain.  At start-up, the plugin automatically determines which element each sample point is inside, and then performs a series of Newton iterations to invert the physical-to-reference space mapping.

\paragraph{Time averaging.}
The \texttt{soln-plugin-tavg} computes the time-average of one or more arbitrary functions.  The functions, which are specified in the configuration file, can be parameterised by both the primitive variables and gradients thereof.  Beyond computing time-averages, the plugin also computes variances, which can be used for the purposes of uncertainty quantification.  Command line support is also included for merging together multiple time-average files.  This is particularly useful in environments where there is a limit on job run time.

\paragraph{Force calculation.}
The \texttt{soln-plugin-fluidforce} can be used to compute the net force on boundaries which, in turn, can be used to obtain aerodynamic quantities such as lift and drag. The plugin breaks out separately the pressure and viscous components of the recorded forces.

\paragraph{Turbulence generation.}
The \texttt{solver-plugin-turbulence} implements the synthetic eddy method of \cite{giangaspero2022synthetic,10.52843/cassyni.249z01}, which allows turbulence to be injected into any portion of the domain. Specifically, isotropic eddies are injected via a source term formulation, where the turbulence intensity and length scale of the eddies can be specified. The implementation is designed to scale efficiently and minimise memory bandwidth requirements. Specifically, by pre-computing and caching element intersections of every injected eddy \emph{before} the simulation starts, the cost of the implementation is able to scale as the number of eddies intersecting an element at a given time---which for sensible grid resolutions will remain small---as opposed to scaling with the overall number of injected eddies, which can be substantial for large domains and small turbulent length scales. Additionally, the implementation also innovates by passing single unsigned 32-bit integer seeds to define multiple characteristics of a given eddy. These are then unrolled by a device side implementation of a PCG random number generator \cite{ONeill2014PCGA} to produce the actual random characteristics of the eddy. This saves memory bandwidth \emph{cf.} pre-computing random eddy characteristics a priori and passing them as an array of (potentially 64-bit) floats.

\paragraph{In-situ visualisation.}
The \texttt{soln-plugin-ascent} provides in-situ visualisation capabilities and is powered by the lightweight Ascent library \cite{larsen2022ascent}. Using the plugin, it is possible to produce complex renderings of the current simulation state without having to write any intermediate files to disk. This enables efficient visualisation of large-scale simulations, for which writing solutions to disk for after-the-fact post-processing is unfeasible.

%% file: figs/fr-mem-aos.tex
\begin{tikzpicture}[scale=1.2]
 \edef\n{18}
 \edef\step{0.2}

    \definecolor{red30}{HTML}{ffb3b8};
    \definecolor{blue30}{HTML}{a6c8ff};

 \foreach \i in {1, ..., \n}
 {
  \pgfmathparse{mod(\i,2) ? "red30" : "blue30"}
  \edef\colour{\pgfmathresult}
  \fill[\colour] (\i*\step - \step,0) rectangle (\i*\step,1);
 }

 \draw[step=\step,gray,thin,line cap=round] (0,0) grid (\step*\n, 1);
\end{tikzpicture}

%% file: figs/fr-mem-soa.tex
\begin{tikzpicture}[scale=1.2]
    \edef\n{18}
    \edef\step{0.2}
 
    \definecolor{red30}{HTML}{ffb3b8};
    \definecolor{blue30}{HTML}{a6c8ff};

    \fill[red30] (0,0) rectangle (0.5*\n*\step,1);
    \fill[blue30] (0.5*\n*\step,0) rectangle (\n*\step,1);

    \draw[step=\step,gray,thin,line cap=round] (0,0) grid (\step*\n, 1);
\end{tikzpicture}

%% file: figs/fr-mem-aosoa.tex
\begin{tikzpicture}[scale=1.2]
    \edef\n{6}
    \edef\k{3}
    \edef\step{0.2}

    \definecolor{red30}{HTML}{ffb3b8};
    \definecolor{blue30}{HTML}{a6c8ff};

    \foreach \i in {1, ..., \n}
    {
        \pgfmathparse{mod(\i,2) ? "red30" : "blue30"}
        \edef\colour{\pgfmathresult}
        \fill[\colour] (\i*\k*\step - \k*\step,0) rectangle (\i*\k*\step,1);
    }

    \draw[step=\step,gray,thin,line cap=round] (0,0) grid (0.2*\k*\n, 1);
\end{tikzpicture}

%% file: community.tex
\section{Developer and User Community}

Over the past decade, an international community of developers and users has grown around PyFR, drawn from across academia and industry. Key to this growth has been the open-source nature of PyFR, which removes many international and inter-institutional barriers to collaborative code development practices. Code development has also been supported at a technical level by hosting the code base in a Git repository on GitHub, which provides a wide range of tooling and helps define best-practice collaborative processes. Also of importance has been maintaining comprehensive and up-to-date documentation using Sphinx, which is auto-deployed to Read the Docs on each release, as well as providing developer and user support via a forum hosted on Discourse. Finally, in 2020, we launched a virtual PyFR seminar series on Cassyni, which comprises invited talks and discussions on a range of topics related to the theory of high-order FR schemes, their implementation in PyFR and their application to industrially relevant flow problems.

\begin{figure*}
    \centering
    {\includegraphics[width=0.5\textwidth]{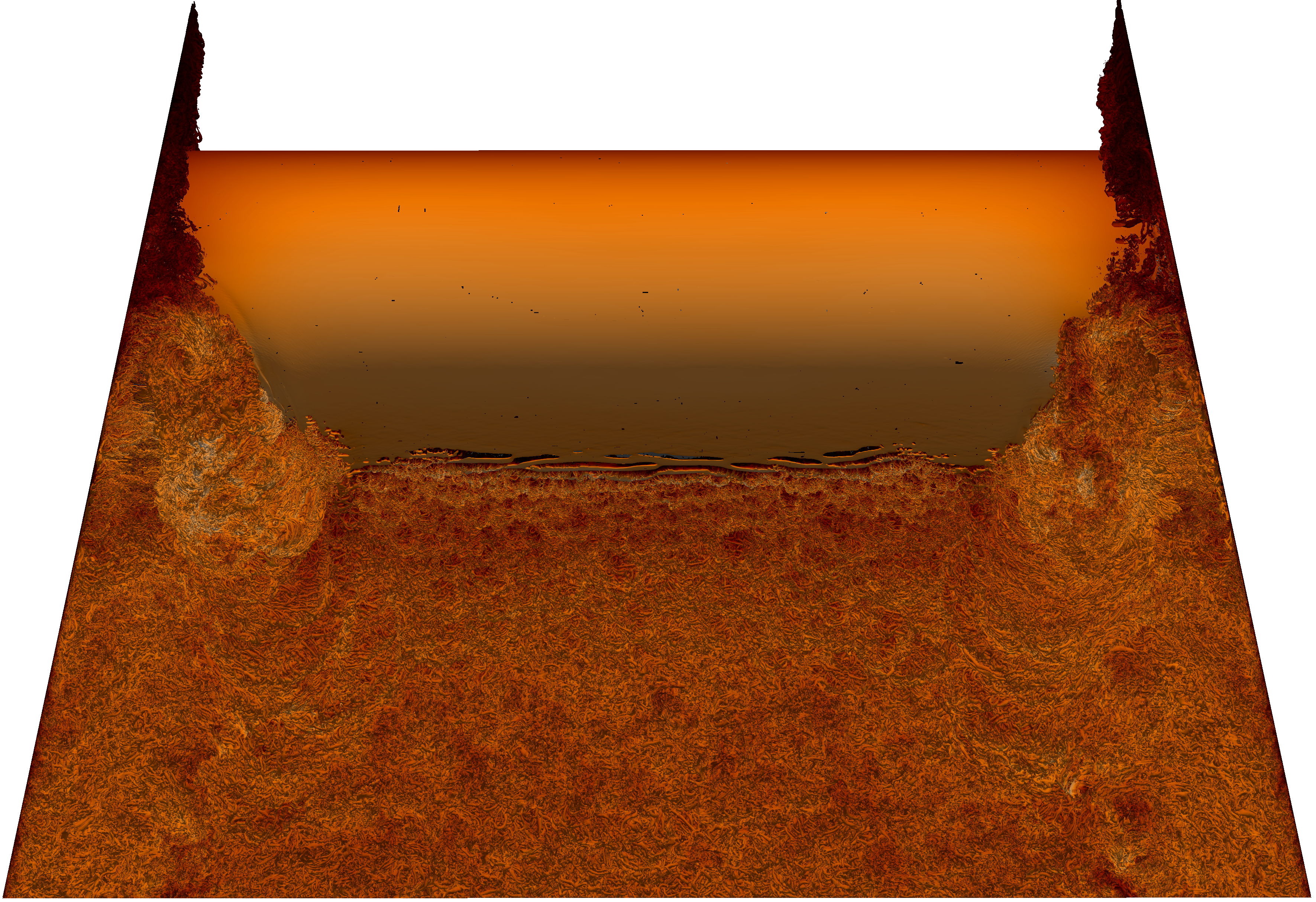}}%
    \caption{Instantaneous snapshot of a Q-criteria iso-surface coloured by velocity magnitude above the suction side of an MTU-T161 low pressure turbine blade obtained using the compressible Navier--Stokes solver in PyFR. Image is from Fig. 12 of Iyer et. al \cite{IYER2021104989}. Copyright Iyer et al. Reused with permission.}
    \label{fig:mtu}
\end{figure*}

\begin{figure*}
    \centering
    {\includegraphics[width=0.4\textwidth]{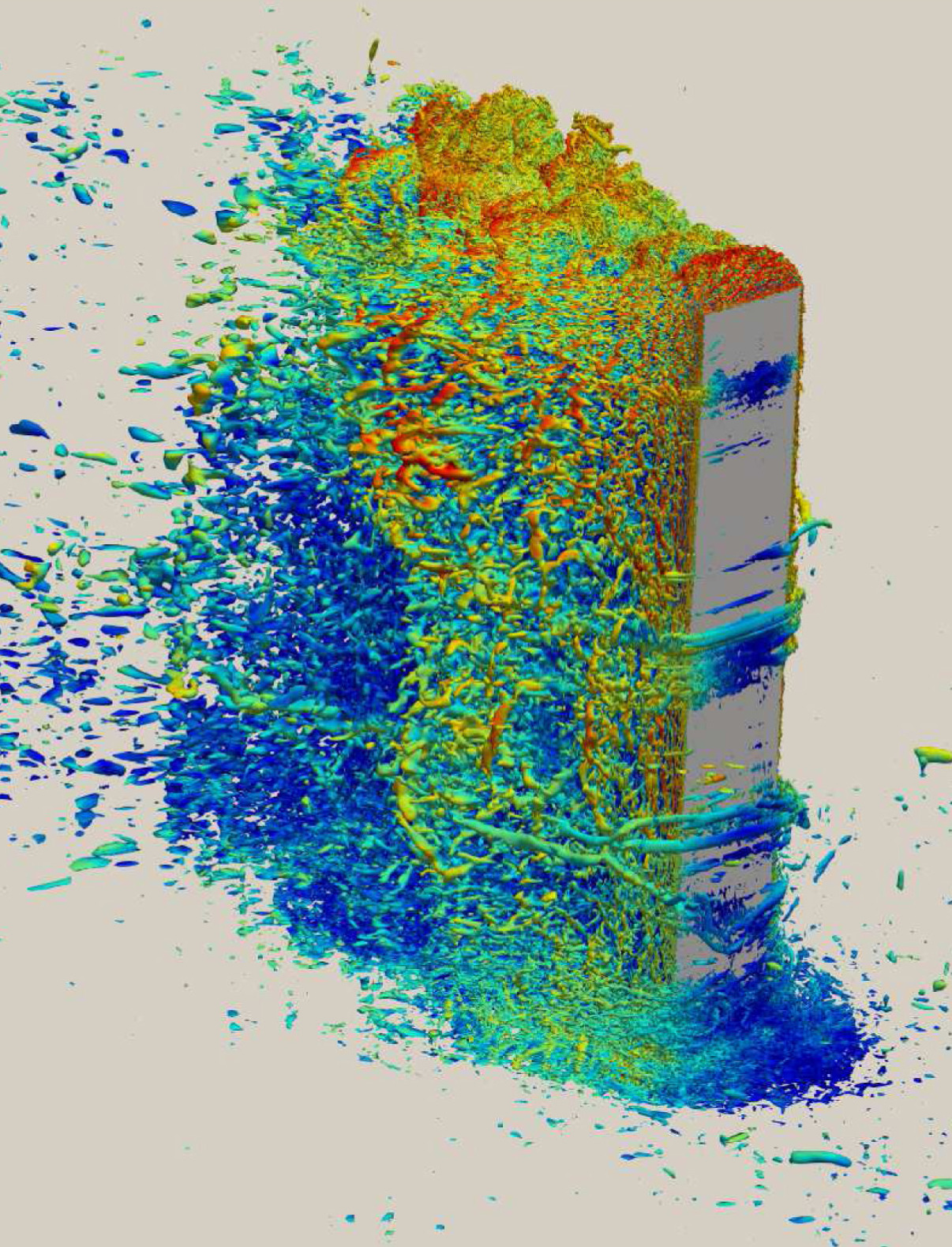}}%
    \caption{Instantaneous snapshot of a Q-criteria iso-surface coloured by velocity magnitude around a model high-rise building obtained using the incompressible Navier--Stokes solver in PyFR. Image is from Fig. 9 of Giangaspero et. al \cite{GIANGASPERO2022105169}. Copyright Giangaspero et al. Reused with permission.}
    \label{fig:arup}
\end{figure*}

\begin{figure*}
    \centering
    {\includegraphics[width=0.5\textwidth]{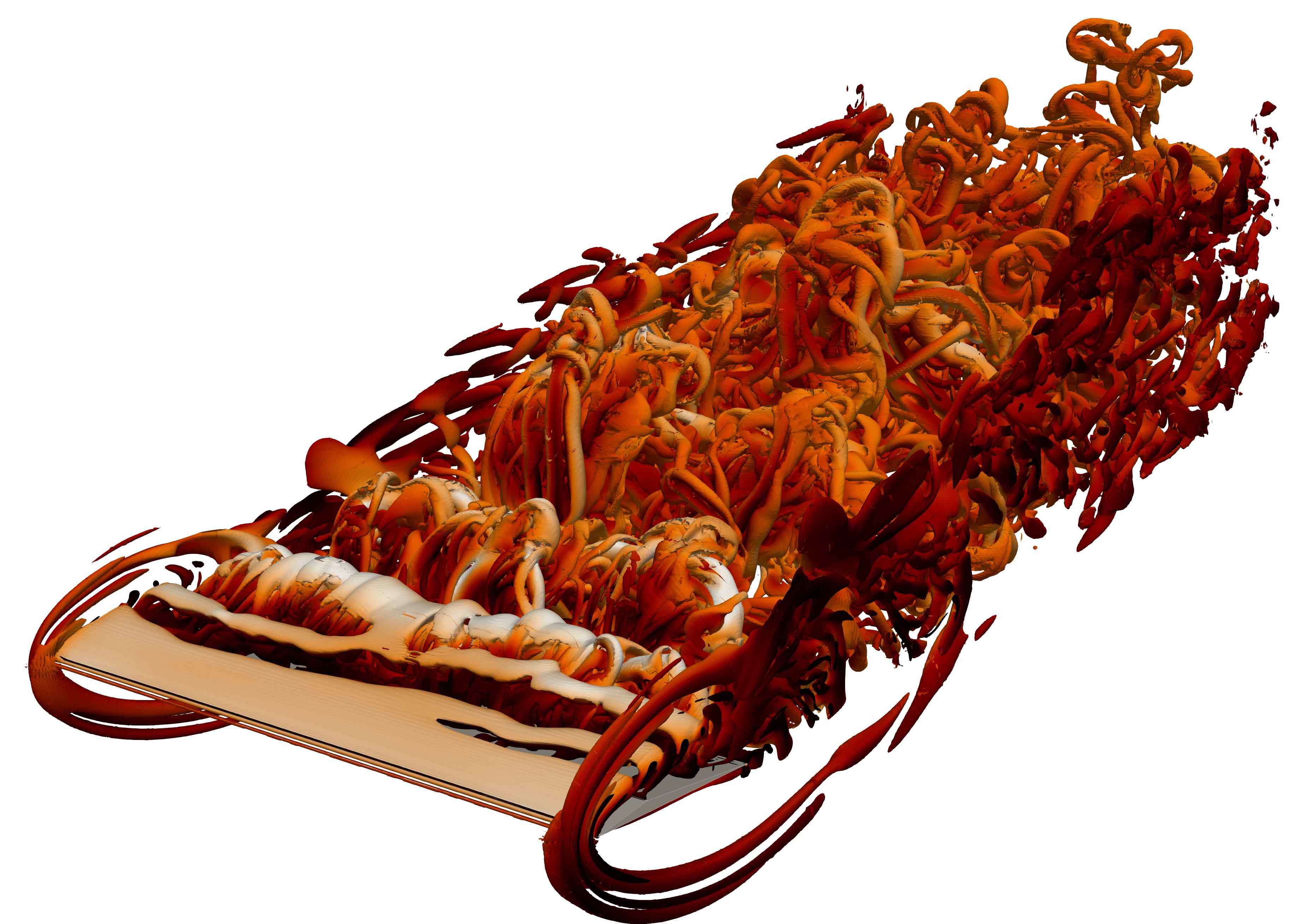}}%
    \caption{Instantaneous snapshot of a Q-criteria iso-surface coloured by velocity magnitude around a triangular aerofoil for a Martian helicopter obtained using the compressible Navier--Stokes solver in PyFR. Image is from Fig. 10 of Caros et. al \cite{doi:10.2514/1.J061454}. Copyright Caros et al. Reused with permission.}
    \label{fig:nasa}
\end{figure*}

Our user base has successfully applied PyFR to a wide range of fundamental, applied, and industrial flow problems, including studies of flow over turbine cascades \cite{10.52843/cassyni.pw9418, IYER2021104989} with MTU Aero Engines (see \cref{fig:mtu}), flow over high-rise buildings \cite{GIANGASPERO2022105169} with Arup (see \cref{fig:arup}), flow over Martian rotorcraft aerofoils \cite{10.52843/cassyni.6r5ry1, 10.52843/47ly7q} with NASA (see \cref{fig:nasa}), flow over supersonic re-entry capsules \cite{doi:10.2514/6.2024-4127} led by NASA, flow over projectiles \cite{10.52843/cassyni.536kkl} led by the Agency for Defense Development in South Korea, flow over wind turbines \cite{10.52843/cassyni.p3fyks, LIANG2024121092}, and flow in thermoacoustic engines \cite{10.52843/cassyni.dlsjz8, BLANC2024122817}, as well as studies of airfoil noise reduction \cite{10.52843/cassyni.br5jn1,10.52843/cassyni.j59tff}, flow control \cite{10.52843/cassyni.nd09lk,10.52843/cassyni.2g7fx6}, wall roughness \cite{10.52843/cassyni.hkqms2}, the Coanda effect \cite{10.52843/cassyni.5yklnq}, surrogate model development \cite{h10.52843/cassyni.s1q5yf, 10.1063/5.0087488} and fundamental aspects of channel flow \cite{10.52843/cassyni.2g7fx6, Iyer_Witherden_Chernyshenko_Vincent_2019}. More recently, PyFR has also been used to enable, for the first time, ILES-based optimisation of turbine cascades \cite{10.52843/cassyni.nqp2sp} and DNS-based optimisation of Martian rotorcraft aerofoils \cite{doi:10.2514/1.J063164}.

%\begin{figure*}
%    \centering
%    {\includegraphics[width=0.5\textwidth]{figs/examples/tarik.jpg}}%
%    \caption{Instantaneous snapshot of density gradient magnitude for a viscous shock tube problem obtained using the compressible Navier--Stokes solver in PyFR. Image is from Fig. 8 of Dzanic et. al \cite{dzanic2022positivity}. Copyright Elsevier Inc. Reused with permission.}
%    \label{fig:tarik}
%\end{figure*}

%% file: frontier.tex
\section{Accuracy, Performance, and Scaling}

\subsection{Accuracy}

To demonstrate the accuracy of PyFR v2.0.3 for high-speed flows, we consider a supersonic Taylor--Green vortex test case at a Mach number of 1.25 and a Reynolds number of $1,600$, which was studied in \citet{Lusher2021} and used to benchmark the accuracy and shock-resolving capabilities of several solvers in \citet{Chapelier2024}. Specifically, we solve the compressible Navier-Stokes equations in a domain $-\pi \leq x,y,z \leq \pi$, subject to the following initial conditions,
\begin{subequations}
    \begin{align}
        u(t=0,\mathbf{x}) &= \sin(x)\cos(y)\cos(z), \\
        v(t=0,\mathbf{x}) &= -\cos(x)\sin(y)\cos(z), \\
        w(t=0,\mathbf{x}) &= 0, \\
        p(t=0, \mathbf{x}) &= p_0 + \frac{1}{16}\left[\cos(2x) + \cos(2y)\right] \left[2 +\cos(2z)\right], \\
        \rho(t=0, \mathbf{x}) &= p(t=0,\mathbf{x})/p_0,
    \end{align}
\end{subequations}
where $p_0$ and the reference dynamic viscosity were selected to achieve the desired Mach and Reynolds numbers based on length, velocity and density scales of unity, and a dynamic viscosity computed using Sutherland's law with a reference temperature of $273$K~\citep{Sutherland1893}. For comparison with the results in \citet{Chapelier2024}, the simulations were performed using computational meshes consisting of $16^3$, $32^3$, $64^3$ and $128^3$ hexahedral elements, with a third-order polynomials approximating the solution within each element (corresponding to $64^3$, $128^3$, $256^3$ and $512^3$ DoFs, respectively). Gauss--Legendre--Lobatto flux and solution points were used, and entropy filtering was employed as a shock capturing approach. 

\begin{figure}[hbtp!]
    \centering
    \subfloat[Enstrophy.]{
    \adjustbox{width=0.5\linewidth,valign=b}{\input{./figs/suptgv_enstrophy}}}
    \hfill
    \subfloat[Schlieren.]{\adjustbox{width=0.4\linewidth,valign=b}{\includegraphics[]{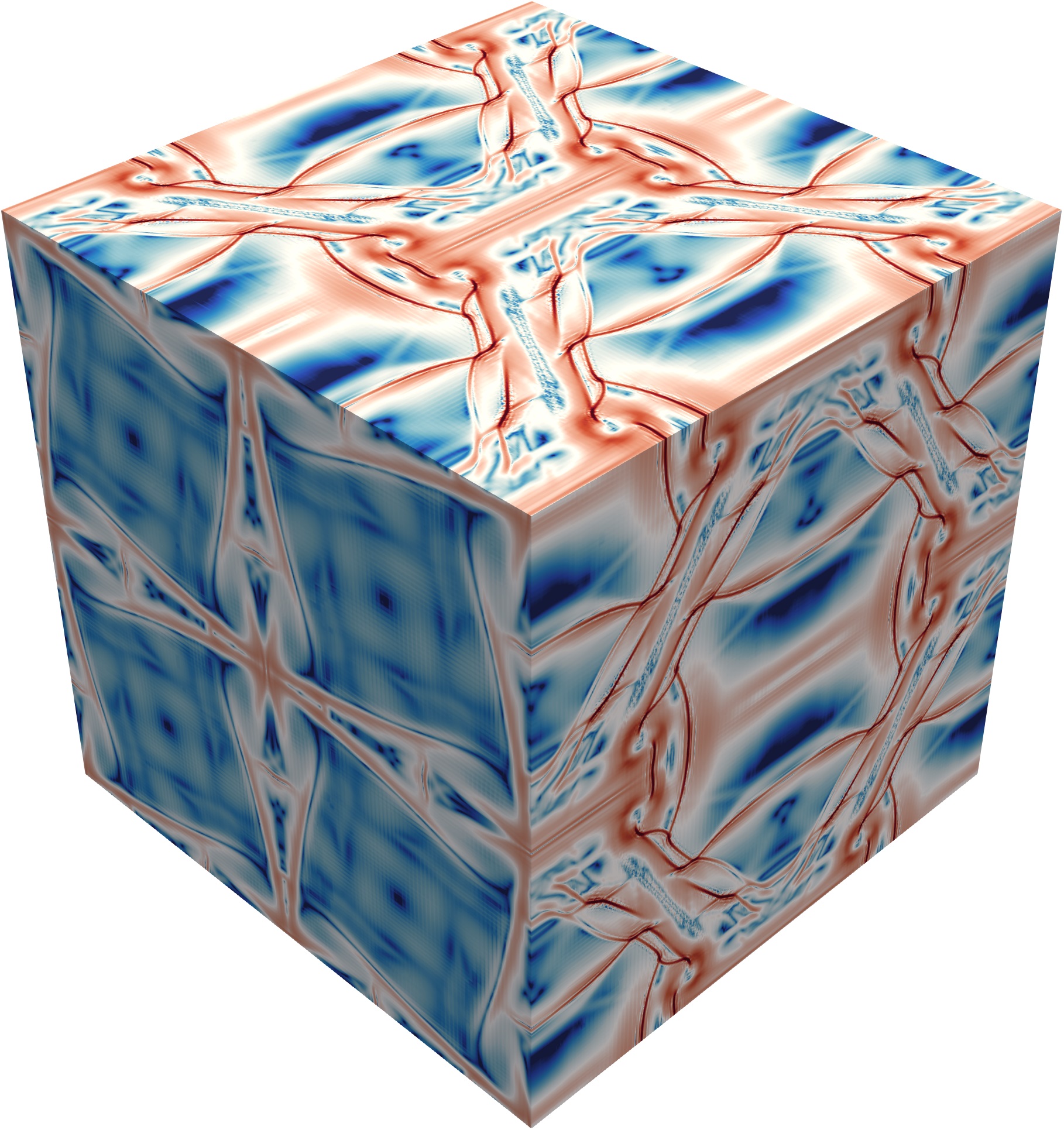}}}
    \caption{
    Plot of enstrophy as a function of time (left) and a Schlieren-type representation of the density gradient norm at $t = 6$ (right) for the supersonic Taylor--Green vortex case computed with $N=128^3$ hexahedral elements ($512^3$ DoFs), along with the reference data of \citet{Chapelier2024}.
    }
    \label{fig:sup_enstrophy}
\end{figure}

\cref{fig:sup_enstrophy} shows solenoidal dissipation (enstrophy), defined as  
\begin{equation}
    \varepsilon_s = \frac{1}{(2 \pi)^3}\int_{-\pi}^{\pi}\int_{-\pi}^{\pi}\int_{-\pi}^{\pi} \mu\,\pmb{\omega}\cdot\pmb{\omega}\;\mathrm{d}x\mathrm{d}y\mathrm{d}z,
\end{equation}
as a function of time, as well as a Schlieren-type representation of the density gradient norm at $t = 6$ for a case computed with $N=128^3$ hexahedral elements ($512^3$ DoFs). The predicted enstrophy profiles are found to be in excellent agreement with the reference data of \citet{Chapelier2024}, computed using a highly resolved ($2048^3$ DoFs) high-order finite difference targeted ENO (TENO) scheme, indicating that PyFR can accurately resolve small-scale turbulent flow structures in supersonic flows. This test case allows for further comparison against the results of various solvers presented in \citet{Chapelier2024}. In particular, we compare to similar discontinuous finite element-type schemes (e.g., Discontinuous Galerkin, spectral difference, etc.) with identical mesh resolution and approximation order, the details of which are summarized in \cref{tab:suptgv-solvers}, as well as the high-order finite difference TENO scheme which was used to compute the reference results.

\begin{table}
  \centering
  \caption{Summary of the solvers and numerical methods used for comparison.  
  }
  \begin{tabular}{r|rrr} \toprule
    Solver & Numerical method  & Order of accuracy & Shock capturing method   \\
    \midrule
    PyFR & Flux Reconstruction & 4 & Entropy filter \\
    CODA~\citep{StefaninVolpiani2024} & Modal Discontinuous Galerkin & 4 & Artificial viscosity \\
    FLEXI~\citep{Krais2021} & Nodal Discontinuous Galerkin  & 4 & Subcell finite volume \\
    SD3D~\citep{Chapelier2016} & Spectral difference & 4 & Artificial viscosity \\
    OpenSBLI~\citep{Lusher2021b} & Finite difference & 6 & Targeted ENO (TENO) \\
    \bottomrule
  \end{tabular}
  \label{tab:suptgv-solvers}
\end{table}

\begin{figure}[hbtp!]
    \centering
    \subfloat[$N_{\text{DoFs}} = 64^3$.]{
    \adjustbox{width=0.45\linewidth,valign=b}{\input{./figs/suptgv_dildis64}}}
    \subfloat[$N_{\text{DoFs}} = 128^3$.]{
    \adjustbox{width=0.45\linewidth,valign=b}{\input{./figs/suptgv_dildis128}}}
    \newline
    \subfloat[$N_{\text{DoFs}} = 256^3$.]{
    \adjustbox{width=0.45\linewidth,valign=b}{\input{./figs/suptgv_dildis256}}}
    \subfloat[$N_{\text{DoFs}} = 512^3$.]{
    \adjustbox{width=0.45\linewidth,valign=b}{\input{./figs/suptgv_dildis512}}}
    \newline
    \caption{
    Dilatational dissipation as a function of time for the supersonic Taylor--Green vortex case computed with varying mesh resolution in comparison to the results of the solvers in \citet{Chapelier2024}.
    }
    \label{fig:sup_dildis}
\end{figure}
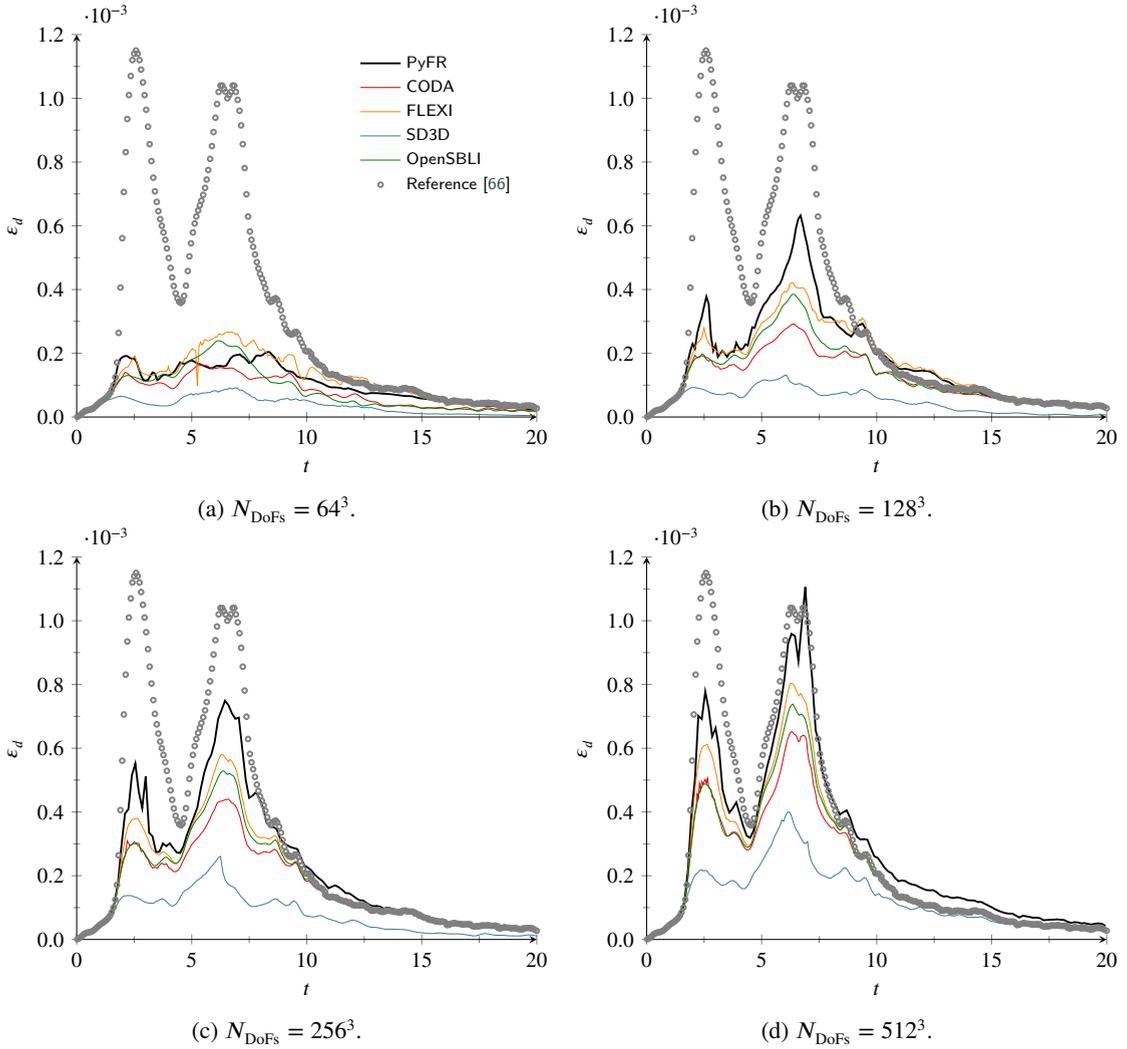

\cref{fig:sup_dildis} shows dilatational  dissipation, defined as  
\begin{equation}
    \varepsilon_d = \frac{4}{3(2 \pi)^3}\int_{-\pi}^{\pi}\int_{-\pi}^{\pi}\int_{-\pi}^{\pi} \mu\, \left (\boldsymbol{\nabla}\cdot \mathbf{u} \right)^2\;\mathrm{d}x\mathrm{d}y\mathrm{d}z,
\end{equation}
as a function of time for cases computed using $64^3$, $128^3$, $256^3$ and $512^3$ DoFs in comparison to the results of the solvers in \citet{Chapelier2024}. The simulations from PyFR generally show less shock dissipation, resulting in dilatational dissipation profiles which are closer to the reference data for a given resolution, indicating that the entropy filtering shock capturing approach does not introduce excessive numerical dissipation and can sharply resolve shock profiles. 

\subsection{Performance and Scaling}

PyFR has previously been used to undertake petascale simulations on a range of the world's largest GPU supercomputers, including Piz Daint at CSCS and Titan at ORNL. Overall performance and strong and weak scaling has been demonstrated previously in this context, and indeed simulations undertaken with PyFR were shortlisted for the Gordon Bell Prize in 2016 \cite{vincent2016towards}, achieving 13.7 DP-PFLOP/s (58 \% of theoretical peak) using 18,000 NVIDIA Tesla K20X GPUs on Titan.
 
\begin{figure}
    \centering
    \begin{tikzpicture}[spy using outlines={rectangle, height=3cm,width=2.3cm, magnification=3, connect spies}]
    \begin{axis}
    [
        axis line style={latex-latex},
        axis y line=left,
        axis x line=left,
        xmode=linear,
        ymode=linear,
        xlabel = {$t$},
        ylabel = {$\varepsilon_s$},
        xmin = 0, xmax = 20,
        ymin = 0, ymax = 0.015,
        minor x tick num=1,
        minor y tick num=1,
        legend cell align={left},
        legend style={font=\scriptsize, at={(0.97, 0.97)},anchor=north east, draw=none},
        %axis line style={draw=none},
        %tick style={draw=none},
        x tick label style={/pgf/number format/.cd, fixed, fixed zerofill, precision=0, /tikz/.cd},
        y tick label style={/pgf/number format/.cd, fixed, fixed zerofill, precision=1, /tikz/.cd},
    ]

        \addplot[color=black, style={very thick}] table[x=t, y=en, col sep=comma]{./figs/integral_p7_m0p08_downscaled.csv};
        \addlegendentry{PyFR}
        
        \addplot[ color=gray, style={thick}, only marks, mark=o, mark options={scale=0.5}, mark repeat = 20, mark phase = 0] table[x=t, y=dkep, col sep=comma, mark=*]{./figs/van_rees_2011_dkep_PSP512.csv};
        \addlegendentry{Reference \cite{VANREES20112794}}
    
    \end{axis}
\end{tikzpicture}
    \caption{Plot of enstrophy as a function of time for the subsonic Taylor--Green vortex case run with PyFR on 512 AMD Instinct MI250X accelerators of Frontier (each with two GCDs), along with the reference data of van Rees et al. \cite{VANREES20112794}.}
    \label{fig:enstrophy}
\end{figure}
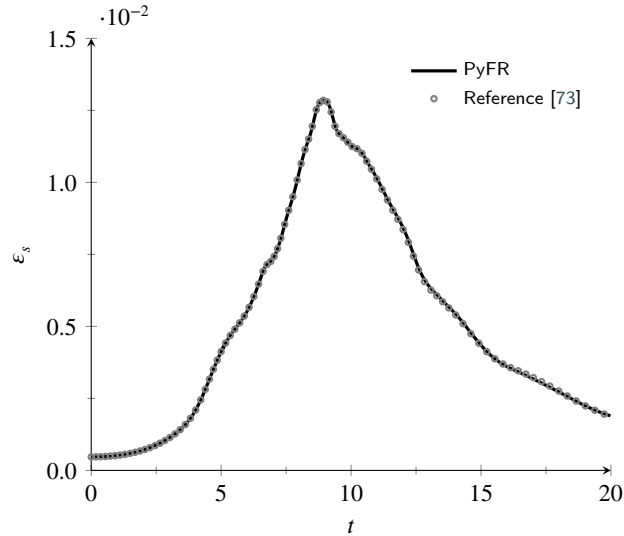

To demonstrate the performance and scaling characteristics of PyFR v2.0.3, we consider a subsonic version of the Taylor--Green vortex test case described above, run with double precision arithmetic on Frontier at ORNL using AMD Instinct MI250X accelerators, and on Alps at CSCS using NVIDIA GH200 GPUs. For this case, the reference pressure was modified to achieve a Mach number of $0.08$, and the dynamic viscosity was set to be constant. The computational mesh consisted of $13,891,500$ tetrahedral elements, with seventh-order solution polynomials used to represent the solution within each element, and alpha-optimised flux and solution points were employed. \cref{fig:enstrophy} plots enstrophy as a function of time for a case run on 512 AMD Instinct MI250X accelerators of Frontier (each with two GCDs). Results are found to be in excellent agreement with the reference data of van Rees et al. \cite{VANREES20112794}.

\cref{tab:ns-strong-scale-frontier} and \cref{tab:ns-strong-scale-alps} present strong scaling of the test case on Frontier and Alps, respectively, where we note that on Frontier PyFR was run without HIP-Aware MPI, whereas on Alps, PyFR was run with CUDA-Aware MPI. In terms of scaling, we observe that at $2048$ ranks, both platforms deliver similar scalability numbers.  However, on account of the superior baseline performance, the NVIDIA system is transferring ${\sim}3.7$ times more data over the interconnect.  Given both systems make use of the Cray Slingshot interconnect, this suggests that the network is \emph{not} the limiting factor on Frontier.  Rather, it is more likely related to our inability to run with HIP-aware MPI on Frontier, and the inability of PyFR to employ native HIP graphs due to unresolved issues in the HIP runtime.

In terms of absolute performance, we note that a single NVIDIA GH200 is ${\sim}3.7$ times faster than one GCD of an AMD MI250X.  A substantial portion of this can be explained by the differences in peak memory bandwidth.  A GH200 uses HBM3 memory with a peak bandwidth of $4\,\text{TiB / s}$, whereas the MI250X uses HBM2e memory with a peak bandwidth per-GCD of $1.6\,\text{TiB / s}$.  This gives a ratio of $2.5$.  Moreover, micro-benchmarks on the MI250X indicate that peak bandwidth is only reliably achieved for kernels with a 1:1 read-to-write ratio.  Outside of this regime, bandwidths closer to ${\sim}1.2\,\text{TiB / s}$ are more commonly observed.  Such a discrepancy is not observed on NVIDIA hardware, however.  Accounting for this gives us a revised performance ratio of ${\sim} 3.3$ which is similar to what is actually observed. The remaining performance differences are likely due to the superior caching setup of the NVIDIA GPU which has---in the absence of shared memory allocations---some $208\,\text{KiB}$ available per SM, whereas AMD only provides $16\,\text{KiB}$ per CU.  Similarly, whereas NVIDIA provide $50\,\text{MiB}$ of shared L2 cache, AMD only provide $8\,\text{MiB}$.  These caches are important for the interface kernels which have an irregular memory access pattern.  

Finally, we can make a comparison between absolute performance almost a decade ago using PyFR v0.2.2 on an NVIDIA K40c GPU \cite{witherden2015heterogeneous} with current absolute performance. Specifically, data from Table 6 of \cite{witherden2015heterogeneous} for a tetrahedrally-dominated mesh with fourth-order solution polynomials in each element gives an absolute performance of $0.122\,\text{GDoF/s}$ per K40c GPU, whereas the $16$ rank case from Table \ref{tab:ns-strong-scale-alps} here gives an absolute performance of $6.004\,\text{GDoF/s}$ per GH200 GPU. This leads to an absolute performance improvement ratio of $49.2$, accounting for the totality of both hardware and software improvements over the period, and where we note the ratio is conservative since use of seventh vs. fourth order solution polynomials necessitates substantially more FLOPs/DoF. This conservative estimate of an almost $50\times$ performance increase over the last decade constitutes a substantial step towards the industrial adoption of scale-resolving simulations.

\begin{table}
  \centering
  \caption{Strong scalability of PyFR on Frontier for the Taylor--Green vortex test case using compressible Navier--Stokes solver on a mesh with $12\times 105^3 = 13,891,500$ tetrahedral elements and seventh-order solution polynomials used to represent the solution within each element.  The speedup is relative to 8 AMD Mi250X accelerators each with two GCDs. HIP-Aware MPI was not employed.}
  \begin{tabular}{r|rrrrrrrr} \toprule
    \# Ranks & 16    & 32    & 64    & 128    & 256   & 512 & 1024 & 2048 \\
    GDoF/s   & 26.21 & 50.87 & 98.70 & 193.37 & 369.10 & 669.80 & 1133.48 & 1537.68 \\
    Speedup   & 1.0 & 1.94 & 3.77 & 7.38 & 14.08 & 25.56 & 43.25 & 58.68 \\
    Efficiency & 1.00 & 0.97 & 0.94 & 0.92 & 0.88 & 0.80 & 0.68 & 0.46 \\
    \bottomrule
  \end{tabular}
  \label{tab:ns-strong-scale-frontier}
\end{table}

\begin{table}
  \centering
  \caption{Strong scalability of PyFR on Alps for the Taylor--Green vortex test case using compressible Navier-Stokes solver on a mesh with $12\times 105^3 = 13,891,500$ tetrahedral elements and seventh-order solution polynomials used to represent the solution within each element.  The speedup is relative to 16 GH200 GPUs. CUDA-Aware MPI was employed.}
  \begin{tabular}{r|rrrrrrrr} \toprule
    \# Ranks & 16    & 32    & 64    & 128    & 256   & 512 & 1024 & 2048 \\
    GDoF/s   & 96.07 & 199.60 & 404.73 & 806.95 & 1599.12 & 2861.56 & 4632.03 & 5753.96 \\
    Speedup   & 1.0 & 2.08 & 4.21 & 8.40 & 16.64 & 29.79 & 48.21 & 59.89 \\
    Efficiency & 1.00 & 1.04 & 1.05 & 1.05 & 1.04 & 0.93 & 0.75	& 0.47 \\
    \bottomrule
  \end{tabular}
  \label{tab:ns-strong-scale-alps}
\end{table}

%% file: figs/suptgv_enstrophy.tex
    \begin{tikzpicture}[spy using outlines={rectangle, height=3cm,width=2.3cm, magnification=3, connect spies}]
    \begin{axis}
    [
        axis lines=left,
        xmode=linear,
        ymode=linear,
        xlabel = {$t$},
        ylabel = {$\varepsilon_s$},
        xmin = 0, xmax = 20,
        ymin = 0, ymax = 0.012,
        minor x tick num=1,
        minor y tick num=1,
        legend cell align={left},
        legend style={font=\scriptsize, at={(0.03, 0.97)},anchor=north west, draw=none},
        %axis line style={draw=none},
        %tick style={draw=none},
        x tick label style={/pgf/number format/.cd, fixed, fixed zerofill, precision=0, /tikz/.cd},
        y tick label style={/pgf/number format/.cd, fixed, fixed zerofill, precision=1, /tikz/.cd},
        scale = 1
    ]

        \addplot[color=black, style={very thick}] table[x=t, y=soldis, col sep=comma]{./figs/data/pyfr_n512.csv};
        \addlegendentry{PyFR}
        
        \addplot[ color=gray, style={thick}, only marks, mark=o, mark options={scale=0.5}, mark repeat = 10, mark phase = 0] table[x=t, y=soldis, col sep=comma, mark=*]{./figs/data/OpenSBLI_TENO6_N2048_M125_Re1600.csv};
        \addlegendentry{Reference~\citep{Chapelier2024}}
        
    \end{axis}
\end{tikzpicture}

%% file: figs/suptgv_dildis64.tex
    \begin{tikzpicture}[spy using outlines={rectangle, height=3cm,width=2.3cm, magnification=3, connect spies}]
    \begin{axis}
    [
        axis lines=left,
        xmode=linear,
        ymode=linear,
        xlabel = {$t$},
        ylabel = {$\varepsilon_d$},
        xmin = 0, xmax = 20,
        ymin = 0, ymax = 0.0012,
        minor x tick num=1,
        minor y tick num=1,
        legend cell align={left},
        legend style={font=\scriptsize, at={(0.97, 0.97)},anchor=north east, draw=none},
        %axis line style={draw=none},
        %tick style={draw=none},
        x tick label style={/pgf/number format/.cd, fixed, fixed zerofill, precision=0, /tikz/.cd},
        y tick label style={/pgf/number format/.cd, fixed, fixed zerofill, precision=1, /tikz/.cd},
        scale = 1
    ]
        
        \addplot[color=black, style={thick}] table[x=t, y=dildis, col sep=comma]{./figs/data/PyFR_N64.csv};
        \addlegendentry{PyFR}

        \addplot[color=red!90!black, style={}] table[x=t, y=dildis, col sep=comma]{./figs/data/CODA_DGp3_N064.csv};
        \addlegendentry{CODA}
        \addplot[color=orange, style={}] table[x=t, y=dildis, col sep=comma]{./figs/data/FLEXI_DGSEM4_N064.csv};
        \addlegendentry{FLEXI}
        \addplot[color=cyan!50!black, style={}] table[x=t, y=dildis, col sep=comma]{./figs/data/SD3D_SD4_N064.csv};
        \addlegendentry{SD3D}
        \addplot[color=green!50!black, style={}] table[x=t, y=dildis, col sep=comma]{./figs/data/OpenSBLI_TENO6_N064.csv};
        \addlegendentry{OpenSBLI}
        
        \addplot[ color=gray, style={thick}, only marks, mark=o, mark options={scale=0.5}, mark repeat = 3, mark phase = 0] table[x=t, y=dildis, col sep=comma, mark=*]{./figs/data/OpenSBLI_TENO6_N2048_M125_Re1600.csv};
        \addlegendentry{Reference~\citep{Chapelier2024}}

    \end{axis}
\end{tikzpicture}

%% file: figs/suptgv_dildis128.tex
    \begin{tikzpicture}[spy using outlines={rectangle, height=3cm,width=2.3cm, magnification=3, connect spies}]
    \begin{axis}
    [
        axis lines=left,
        xmode=linear,
        ymode=linear,
        xlabel = {$t$},
        ylabel = {$\varepsilon_d$},
        xmin = 0, xmax = 20,
        ymin = 0, ymax = 0.0012,
        minor x tick num=1,
        minor y tick num=1,
        legend cell align={left},
        legend style={font=\scriptsize, at={(0.97, 0.97)},anchor=north east, draw=none},
        %axis line style={draw=none},
        %tick style={draw=none},
        x tick label style={/pgf/number format/.cd, fixed, fixed zerofill, precision=0, /tikz/.cd},
        y tick label style={/pgf/number format/.cd, fixed, fixed zerofill, precision=1, /tikz/.cd},
        scale = 1
    ]
        
        \addplot[color=black, style={thick}] table[x=t, y=dildis, col sep=comma]{./figs/data/PyFR_N128.csv};
        % \addlegendentry{PyFR}

        \addplot[color=red!90!black, style={}] table[x=t, y=dildis, col sep=comma]{./figs/data/CODA_DGp3_N128.csv};
        % \addlegendentry{CODA}
        \addplot[color=orange, style={}] table[x=t, y=dildis, col sep=comma]{./figs/data/FLEXI_DGSEM4_N128.csv};
        % \addlegendentry{FLEXI}
        \addplot[color=cyan!50!black, style={}] table[x=t, y=dildis, col sep=comma]{./figs/data/SD3D_SD4_N128.csv};
        % \addlegendentry{SD3D}
        \addplot[color=green!50!black, style={}] table[x=t, y=dildis, col sep=comma]{./figs/data/OpenSBLI_TENO6_N128.csv};
        % \addlegendentry{OpenSBLI}
        
        \addplot[ color=gray, style={thick}, only marks, mark=o, mark options={scale=0.5}, mark repeat = 3, mark phase = 0] table[x=t, y=dildis, col sep=comma, mark=*]{./figs/data/OpenSBLI_TENO6_N2048_M125_Re1600.csv};
        % \addlegendentry{Reference~\citep{Chapelier2024}}
        
    \end{axis}
\end{tikzpicture}

%% file: figs/suptgv_dildis256.tex
    \begin{tikzpicture}[spy using outlines={rectangle, height=3cm,width=2.3cm, magnification=3, connect spies}]
    \begin{axis}
    [
        axis lines=left,
        xmode=linear,
        ymode=linear,
        xlabel = {$t$},
        ylabel = {$\varepsilon_d$},
        xmin = 0, xmax = 20,
        ymin = 0, ymax = 0.0012,
        minor x tick num=1,
        minor y tick num=1,
        legend cell align={left},
        legend style={font=\scriptsize, at={(0.97, 0.97)},anchor=north east, draw=none},
        %axis line style={draw=none},
        %tick style={draw=none},
        x tick label style={/pgf/number format/.cd, fixed, fixed zerofill, precision=0, /tikz/.cd},
        y tick label style={/pgf/number format/.cd, fixed, fixed zerofill, precision=1, /tikz/.cd},
        scale = 1
    ]
        
        \addplot[color=black, style={thick}] table[x=t, y=dildis, col sep=comma]{./figs/data/PyFR_N256.csv};
        % \addlegendentry{PyFR}

        \addplot[color=red!90!black, style={}] table[x=t, y=dildis, col sep=comma]{./figs/data/CODA_DGp3_N256.csv};
        % \addlegendentry{CODA}
        \addplot[color=orange, style={}] table[x=t, y=dildis, col sep=comma]{./figs/data/FLEXI_DGSEM4_N256.csv};
        % \addlegendentry{FLEXI}
        \addplot[color=cyan!50!black, style={}] table[x=t, y=dildis, col sep=comma]{./figs/data/SD3D_SD4_N256.csv};
        % \addlegendentry{SD3D}
        \addplot[color=green!50!black, style={}] table[x=t, y=dildis, col sep=comma]{./figs/data/OpenSBLI_TENO6_N256.csv};
        % \addlegendentry{OpenSBLI}
        
        \addplot[ color=gray, style={thick}, only marks, mark=o, mark options={scale=0.5}, mark repeat = 3, mark phase = 0] table[x=t, y=dildis, col sep=comma, mark=*]{./figs/data/OpenSBLI_TENO6_N2048_M125_Re1600.csv};
        % \addlegendentry{Reference~\citep{Chapelier2024}}
        
    \end{axis}
\end{tikzpicture}

%% file: figs/suptgv_dildis512.tex
    \begin{tikzpicture}[spy using outlines={rectangle, height=3cm,width=2.3cm, magnification=3, connect spies}]
    \begin{axis}
    [
        axis lines=left,
        xmode=linear,
        ymode=linear,
        xlabel = {$t$},
        ylabel = {$\varepsilon_d$},
        xmin = 0, xmax = 20,
        ymin = 0, ymax = 0.0012,
        minor x tick num=1,
        minor y tick num=1,
        legend cell align={left},
        legend style={font=\scriptsize, at={(0.97, 0.97)},anchor=north east, draw=none},
        %axis line style={draw=none},
        %tick style={draw=none},
        x tick label style={/pgf/number format/.cd, fixed, fixed zerofill, precision=0, /tikz/.cd},
        y tick label style={/pgf/number format/.cd, fixed, fixed zerofill, precision=1, /tikz/.cd},
        scale = 1
    ]
        
        \addplot[color=black, style={thick}] table[x=t, y=dildis, col sep=comma]{./figs/data/PyFR_N512.csv};
        % \addlegendentry{PyFR}

        \addplot[color=red!90!black, style={}] table[x=t, y=dildis, col sep=comma]{./figs/data/CODA_DGp3_N512.csv};
        % \addlegendentry{CODA}
        \addplot[color=orange, style={}] table[x=t, y=dildis, col sep=comma]{./figs/data/FLEXI_DGSEM4_N512.csv};
        % \addlegendentry{FLEXI}
        \addplot[color=cyan!50!black, style={}] table[x=t, y=dildis, col sep=comma]{./figs/data/SD3D_SD4_N512.csv};
        % \addlegendentry{SD3D}
        \addplot[color=green!50!black, style={}] table[x=t, y=dildis, col sep=comma]{./figs/data/OpenSBLI_TENO6_N512.csv};
        % \addlegendentry{OpenSBLI}
        
        \addplot[ color=gray, style={thick}, only marks, mark=o, mark options={scale=0.5}, mark repeat = 3, mark phase = 0] table[x=t, y=dildis, col sep=comma, mark=*]{./figs/data/OpenSBLI_TENO6_N2048_M125_Re1600.csv};
        % \addlegendentry{Reference~\citep{Chapelier2024}}

    \end{axis}
\end{tikzpicture}

%% file: conclusions.tex
\section{Conclusions}

Since the initial release of PyFR v0.1.0 in 2013 \cite{witherden2014pyfr}, a range of new capabilities have been added to the framework, with a view to enabling industrial adoption. In this work, we have provided details of these enhancements as released in PyFR v2.0.3, including improvements to cross-platform performance (new backends, extensions of the DSL, new matrix multiplication providers, improvements to the data layout, use of task graphs) and improvements to numerical stability (modal filtering, anti-aliasing, artificial viscosity, entropy filtering), as well as the addition of prismatic, tetrahedral and pyramid shaped elements, improved domain decomposition support for mixed element grids, improved handling of curved element meshes, the addition of an adaptive time-stepping capability, the addition of incompressible Euler and Navier-Stokes solvers, improvements to file formats and the development of a plugin architecture. We have also explained efforts to grow an engaged developer and user community and provided a range of examples that demonstrate how our user base is applying PyFR to solve a wide range of fundamental, applied and industrial flow problems. Finally, we have demonstrated the accuracy of PyFR v2.0.3 for a supersonic Taylor-Green vortex case, with shocks and turbulence, and provided latest performance and scaling results on up to 1024 AMD Instinct MI250X accelerators of Frontier at ORNL (each with two GCDs) and up to 2048 Nvidia GH200 GPUs of Alps at CSCS. We note that absolute performance of PyFR accounting for the totality of both hardware and software improvements has, conservatively, increased by almost $50\times$ over the last decade.

%% file: acknowledgements.tex
\section*{Acknowledgements}

FDW would like to acknowledge support from the Air Force Office of Scientific Research via grants FA9550-21-1-0190 (``Enabling next-generation heterogeneous computing for massively parallel high-order compressible CFD'') and FA9550-23-1-0232 (``Next Generation High-Order Methods for Multi-Physics Multi-Scale Problems'') under the direction of Dr. Fariba Fahroo. PEV would like to acknowledge support from the Engineering and Physical Sciences Research Council via awards EP/K027379/1, EP/R030340/1, EP/R029423/1, EP/L000407/1, the European Commission via awards 635962 and 814837, the Leverhulme Trust via a Philip Leverhulme Prize, and Innovate UK via award EP/M50676X/1, and BAE Systems Submarines. BCV acknowledges support from the Natural Sciences and Engineering Research Council of Canada (NSERC) under grants RGPAS-2017-507988 and RGPIN-2017-06773. TD would like to thank Jean-Baptiste Chapelier and collaborators for providing comparison data for simulations in this work. An award of computer time was provided by the Innovative and Novel Computational Impact on Theory and Experiment (INCITE) program. This research used resources of the Oak Ridge Leadership Computing Facility at the Oak Ridge National Laboratory, which is supported by the Office of Science of the U.S. Department of Energy under Contract DE-AC05-00OR22725. This work also used computational resources of TSUBAME through the HPCI System Research Project (Project ID: hp190061), was supported by JST FOREST Program (Grant Number JPMJFR2124) and JSPS KAKENHI (Grant Number 24K01074), and was partially performed under the auspices of the U.S. Department of Energy by Lawrence Livermore National Laboratory under contract DE--AC52--07NA27344.

%% file: main.bbl
\begin{thebibliography}{73}
\providecommand{\natexlab}[1]{#1}
\providecommand{\url}[1]{\texttt{#1}}
\expandafter\ifx\csname urlstyle\endcsname\relax
  \providecommand{\doi}[1]{doi: #1}\else
  \providecommand{\doi}{doi: \begingroup \urlstyle{rm}\Url}\fi

\bibitem[Huynh(2007)]{Huynh2007}
H.~T. Huynh.
\newblock A flux reconstruction approach to high-order schemes including
  discontinuous galerkin methods.
\newblock In \emph{18th AIAA Computational Fluid Dynamics Conference}. American
  Institute of Aeronautics and Astronautics, June 2007.
\newblock \doi{10.2514/6.2007-4079}.

\bibitem[Cantwell et~al.(2015)Cantwell, Moxey, Comerford, Bolis, Rocco,
  Mengaldo, De~Grazia, Yakovlev, Lombard, Ekelschot,
  et~al.]{cantwell2015nektar++}
C.~D. Cantwell, D.~Moxey, A.~Comerford, A.~Bolis, G.~Rocco, G.~Mengaldo,
  D.~De~Grazia, S.~Yakovlev, J.-E. Lombard, D.~Ekelschot, et~al.
\newblock Nektar++: An open-source spectral/hp element framework.
\newblock \emph{Computer Physics Communications}, 192:\penalty0 205--219, 2015.

\bibitem[Moxey et~al.(2020)Moxey, Cantwell, Bao, Cassinelli, Castiglioni, Chun,
  Juda, Kazemi, Lackhove, Marcon, et~al.]{moxey2020nektar++}
D.~Moxey, C.~D. Cantwell, Y.~Bao, A.~Cassinelli, G.~Castiglioni, S.~Chun,
  E.~Juda, E.~Kazemi, K.~Lackhove, J.~Marcon, et~al.
\newblock Nektar++: Enhancing the capability and application of high-fidelity
  spectral/hp element methods.
\newblock \emph{Computer Physics Communications}, 249:\penalty0 107110, 2020.

\bibitem[Wang et~al.(2017)Wang, Li, Jia, Laskowski, Kopriva, Paliath, and
  Bhaskaran]{WANG2017579}
Z.J. Wang, Y.~Li, F.~Jia, G.M. Laskowski, J.~Kopriva, U.~Paliath, and
  R.~Bhaskaran.
\newblock Towards industrial large eddy simulation using the {FR/CPR} method.
\newblock \emph{Computers \& Fluids}, 156:\penalty0 579--589, 2017.

\bibitem[Sandberg(2008)]{sandberg2008development}
R.~D. Sandberg.
\newblock Development of a new compressible navier-stokes solver for numerical
  simulations of flows in turbomachinery.
\newblock \emph{Progress report for HPC Europa++ Transnational Access Project},
  1264, 2008.

\bibitem[Bres et~al.(2018)Bres, Bose, Emory, Ham, Schmidt, Rigas, and
  Colonius]{bres2018large}
G.~A. Bres, S.~T. Bose, M.~Emory, F.~E. Ham, O.~T. Schmidt, G.~Rigas, and
  T.~Colonius.
\newblock Large-eddy simulations of co-annular turbulent jet using a
  voronoi-based mesh generation framework.
\newblock In \emph{2018 AIAA/CEAS Aeroacoustics Conference}, page 3302, 2018.

\bibitem[Goc et~al.(2021)Goc, Lehmkuhl, Park, Bose, and Moin]{goc2021large}
K.~A. Goc, O.~Lehmkuhl, G.~I. Park, S.~T. Bose, and P.~Moin.
\newblock Large eddy simulation of aircraft at affordable cost: a milestone in
  computational fluid dynamics.
\newblock \emph{Flow}, 1:\penalty0 E14, 2021.

\bibitem[Witherden et~al.(2014)Witherden, Farrington, and
  Vincent]{witherden2014pyfr}
F.~D. Witherden, A.~M. Farrington, and P.~E. Vincent.
\newblock {PyFR}: An open source framework for solving advection--diffusion
  type problems on streaming architectures using the flux reconstruction
  approach.
\newblock \emph{Computer Physics Communications}, 185\penalty0 (11):\penalty0
  3028--3040, 2014.

\bibitem[Karniadakis and Sherwin(2005)]{karniadakis2005spectral}
G.~Karniadakis and S.~J. Sherwin.
\newblock \emph{Spectral/hp element methods for computational fluid dynamics}.
\newblock Oxford University Press, USA, 2005.

\bibitem[Hesthaven and Warburton(2007)]{hesthaven2007nodal}
J.~S. Hesthaven and T.~Warburton.
\newblock \emph{Nodal discontinuous {Galerkin} methods: algorithms, analysis,
  and applications}.
\newblock Springer Science \& Business Media, 2007.

\bibitem[Wozniak et~al.(2016)Wozniak, Witherden, Russell, Vincent, and
  Kelly]{wozniak2016gimmik}
B.~D. Wozniak, F.~D. Witherden, F.~P. Russell, P.~E. Vincent, and P.~H.~J.
  Kelly.
\newblock {GiMMiK}—generating bespoke matrix multiplication kernels for
  accelerators: Application to high-order computational fluid dynamics.
\newblock \emph{Computer Physics Communications}, 202:\penalty0 12--22, 2016.

\bibitem[Heinecke et~al.(2016)Heinecke, Henry, Hutchinson, and
  Pabst]{heinecke2016libxsmm}
A.~Heinecke, G.~Henry, M.~Hutchinson, and H.~Pabst.
\newblock {LIBXSMM}: accelerating small matrix multiplications by runtime code
  generation.
\newblock In \emph{SC'16: Proceedings of the International Conference for High
  Performance Computing, Networking, Storage and Analysis}, pages 981--991.
  IEEE, 2016.

\bibitem[Akkurt et~al.(2022)Akkurt, Witherden, and Vincent]{akkurt2022cache}
S.~Akkurt, F.~D. Witherden, and P.~E. Vincent.
\newblock Cache blocking strategies applied to flux reconstruction.
\newblock \emph{Computer Physics Communications}, 271:\penalty0 108193, 2022.

\bibitem[Akkurt(2021)]{10.52843/gpnpwx}
S.~Akkurt.
\newblock {Cache Blocking Strategies Applied to Flux Reconstruction}.
\newblock \emph{PyFR Seminar Series}, 2021.
\newblock \doi{10.52843/gpnpwx}.

\bibitem[Dalc{\'\i}n et~al.(2005)Dalc{\'\i}n, Paz, and Storti]{dalcin2005mpi}
L.~Dalc{\'\i}n, R.~Paz, and M.~Storti.
\newblock {MPI} for {Python}.
\newblock \emph{Journal of Parallel and Distributed Computing}, 65\penalty0
  (9):\penalty0 1108--1115, 2005.

\bibitem[Dalc{\'\i}n et~al.(2008)Dalc{\'\i}n, Paz, Storti, and
  D’El{\'\i}a]{dalcin2008mpi}
L.~Dalc{\'\i}n, R.~Paz, M.~Storti, and J.~D’El{\'\i}a.
\newblock {MPI} for {Python}: Performance improvements and {MPI-2} extensions.
\newblock \emph{Journal of Parallel and Distributed Computing}, 68\penalty0
  (5):\penalty0 655--662, 2008.

\bibitem[Witherden et~al.(2015)Witherden, Vermeire, and
  Vincent]{witherden2015heterogeneous}
F.~D. Witherden, B.~C. Vermeire, and P.~E. Vincent.
\newblock Heterogeneous computing on mixed unstructured grids with {PyFR}.
\newblock \emph{Computers \& Fluids}, 120:\penalty0 173--186, 2015.

\bibitem[Mishra et~al.(2023)Mishra, Witherden, Chakravorty, Perez, and
  Dang]{mishra2023scaling}
S.~Mishra, F.~D. Witherden, D.~Chakravorty, L.~Perez, and F.~Dang.
\newblock Scaling study of flow simulations on composable cyberinfrastructure.
\newblock In \emph{Practice and Experience in Advanced Research Computing},
  pages 221--225. 2023.

\bibitem[Jameson et~al.(2012)Jameson, Vincent, and Castonguay]{jameson2012non}
A.~Jameson, P.~E. Vincent, and P.~Castonguay.
\newblock On the non-linear stability of flux reconstruction schemes.
\newblock \emph{Journal of Scientific Computing}, 50:\penalty0 434--445, 2012.

\bibitem[Witherden and Vincent(2015)]{witherden2015identification}
F.~D. Witherden and P.~E. Vincent.
\newblock On the identification of symmetric quadrature rules for finite
  element methods.
\newblock \emph{Computers \& Mathematics with Applications}, 69\penalty0
  (10):\penalty0 1232--1241, 2015.

\bibitem[Park et~al.(2017)Park, Witherden, and Vincent]{park2017high}
J.~S Park, F.~D Witherden, and P.~E Vincent.
\newblock High-order implicit large-eddy simulations of flow over a {NACA0021}
  aerofoil.
\newblock \emph{AIAA journal}, 55\penalty0 (7):\penalty0 2186--2197, 2017.

\bibitem[Persson and Peraire(2006)]{persson2006sub}
P.-O. Persson and J.~Peraire.
\newblock Sub-cell shock capturing for discontinuous {Galerkin} methods.
\newblock In \emph{44th AIAA aerospace sciences meeting and exhibit}, page 112,
  2006.

\bibitem[Dzanic and Witherden(2022)]{dzanic2022positivity}
T.~Dzanic and F.~D. Witherden.
\newblock Positivity-preserving entropy-based adaptive filtering for
  discontinuous spectral element methods.
\newblock \emph{Journal of Computational Physics}, 468:\penalty0 111501, 2022.

\bibitem[Dzanic and Witherden(2023)]{dzanic2023positivity}
T.~Dzanic and F.~D. Witherden.
\newblock Positivity-preserving entropy filtering for the ideal
  magnetohydrodynamics equations.
\newblock \emph{Computers \& Fluids}, 266:\penalty0 106056, 2023.

\bibitem[Dzanic(2022)]{10.52843/cassyni.pvy6c0}
T.~Dzanic.
\newblock {Positivity-preserving entropy-based adaptive filtering for shock
  capturing}.
\newblock \emph{PyFR Seminar Series}, 2022.
\newblock \doi{10.52843/cassyni.pvy6c0}.

\bibitem[Trojak and Dzanic(2024)]{Trojak2024}
Will Trojak and Tarik Dzanic.
\newblock Positivity-preserving discontinuous spectral element methods for
  compressible multi-species flows.
\newblock \emph{Computers \& Fluids}, 280:\penalty0 106343, August 2024.
\newblock \doi{10.1016/j.compfluid.2024.106343}.

\bibitem[Karypis and Kumar(1997)]{karypis1997metis}
G.~Karypis and V.~Kumar.
\newblock {METIS}: A software package for partitioning unstructured graphs,
  partitioning meshes, and computing fill-reducing orderings of sparse
  matrices.
\newblock 1997.

\bibitem[Chevalier and Pellegrini(2008)]{chevalier2008pt}
C{\'e}dric Chevalier and Fran{\c{c}}ois Pellegrini.
\newblock {PT-Scotch}: A tool for efficient parallel graph ordering.
\newblock \emph{Parallel computing}, 34\penalty0 (6-8):\penalty0 318--331,
  2008.

\bibitem[Kopriva(2006)]{kopriva2006metric}
D.~A. Kopriva.
\newblock Metric identities and the discontinuous spectral element method on
  curvilinear meshes.
\newblock \emph{Journal of Scientific Computing}, 26:\penalty0 301--327, 2006.

\bibitem[Abe et~al.(2015)Abe, Haga, Nonomura, and Fujii]{abe2015freestream}
Y.~Abe, T.~Haga, T.~Nonomura, and K.~Fujii.
\newblock On the freestream preservation of high-order conservative
  flux-reconstruction schemes.
\newblock \emph{Journal of Computational Physics}, 281:\penalty0 28--54, 2015.

\bibitem[Hairer et~al.(1993)Hairer, N{\o}rsett, and Wanner]{heirer1993solving}
E.~Hairer, S.~P. N{\o}rsett, and G.~Wanner.
\newblock \emph{Solving Ordinary Differential Equations I: Nonstiff Problems}.
\newblock Springer, Berlin, 2 edition, 1993.

\bibitem[Chorin(1997)]{chorin1997numerical}
A.~J. Chorin.
\newblock A numerical method for solving incompressible viscous flow problems.
\newblock \emph{Journal of computational physics}, 135\penalty0 (2):\penalty0
  118--125, 1997.

\bibitem[Jameson(1991)]{jameson1991time}
A.~Jameson.
\newblock Time dependent calculations using multigrid, with applications to
  unsteady flows past airfoils and wings.
\newblock In \emph{10th Computational fluid dynamics conference}, page 1596,
  1991.

\bibitem[Loppi et~al.(2018)Loppi, Witherden, Jameson, and
  Vincent]{loppi2018high}
N.~A. Loppi, F.~D. Witherden, A.~Jameson, and P.~E. Vincent.
\newblock A high-order cross-platform incompressible {Navier--Stokes} solver
  via artificial compressibility with application to a turbulent jet.
\newblock \emph{Computer Physics Communications}, 233:\penalty0 193--205, 2018.

\bibitem[Loppi et~al.(2019)Loppi, Witherden, Jameson, and
  Vincent]{loppi2019locally}
N.~A. Loppi, F.~D. Witherden, A.~Jameson, and P.~E. Vincent.
\newblock Locally adaptive pseudo-time stepping for high-order flux
  reconstruction.
\newblock \emph{Journal of Computational Physics}, 399:\penalty0 108913, 2019.

\bibitem[Folk et~al.(2011)Folk, Heber, Koziol, Pourmal, and
  Robinson]{folk2011overview}
M.~Folk, G.~Heber, Q.~Koziol, E.~Pourmal, and D.~Robinson.
\newblock An overview of the {HDF5} technology suite and its applications.
\newblock In \emph{Proceedings of the EDBT/ICDT 2011 workshop on array
  databases}, pages 36--47, 2011.

\bibitem[Collette(2013)]{collette2013python}
A.~Collette.
\newblock \emph{Python and {HDF5}: unlocking scientific data}.
\newblock O'Reilly Media, Inc., 2013.

\bibitem[Giangaspero et~al.(2022{\natexlab{a}})Giangaspero, Witherden, and
  Vincent]{giangaspero2022synthetic}
G.~Giangaspero, F.~D. Witherden, and P.~E. Vincent.
\newblock Synthetic turbulence generation for high-order scale-resolving
  simulations on unstructured grids.
\newblock \emph{AIAA Journal}, 60\penalty0 (2):\penalty0 1032--1051,
  2022{\natexlab{a}}.

\bibitem[Giangaspero(2021)]{10.52843/cassyni.249z01}
G.~Giangaspero.
\newblock {Synthetic Turbulence Generation in PyFR}.
\newblock \emph{PyFR Seminar Series}, 2021.
\newblock \doi{10.52843/cassyni.249z01}.

\bibitem[O'Neill(2014)]{ONeill2014PCGA}
Melissa~E. O'Neill.
\newblock Pcg : A family of simple fast space-efficient statistically good
  algorithms for random number generation.
\newblock 2014.
\newblock URL \url{https://api.semanticscholar.org/CorpusID:3489282}.

\bibitem[Larsen et~al.(2022)Larsen, Brugger, Childs, and
  Harrison]{larsen2022ascent}
M.~Larsen, E.~Brugger, H.~Childs, and C.~Harrison.
\newblock Ascent: A flyweight in situ library for exascale simulations.
\newblock In \emph{In Situ Visualization for Computational Science}, pages
  255--279. Springer, 2022.

\bibitem[Afshar(2021)]{10.52843/cassyni.pw9418}
N.~F. Afshar.
\newblock {High-Order Implicit Large Eddy Simulation of Flow over a
  Low-Reynolds Turbine Cascade}.
\newblock \emph{PyFR Seminar Series}, 2021.
\newblock \doi{10.52843/cassyni.pw9418}.

\bibitem[Iyer et~al.(2021)Iyer, Abe, Vermeire, Bechlars, Baier, Jameson,
  Witherden, and Vincent]{IYER2021104989}
A.S. Iyer, Y.~Abe, B.C. Vermeire, P.~Bechlars, R.D. Baier, A.~Jameson, F.D.
  Witherden, and P.E. Vincent.
\newblock High-order accurate direct numerical simulation of flow over a
  mtu-t161 low pressure turbine blade.
\newblock \emph{Computers \& Fluids}, 226:\penalty0 104989, 2021.

\bibitem[Giangaspero et~al.(2022{\natexlab{b}})Giangaspero, Amerio, Downie,
  Zasso, and Vincent]{GIANGASPERO2022105169}
Giorgio Giangaspero, Luca Amerio, Steven Downie, Alberto Zasso, and Peter
  Vincent.
\newblock High-order scale-resolving simulations of extreme wind loads on a
  model high-rise building.
\newblock \emph{Journal of Wind Engineering and Industrial Aerodynamics},
  230:\penalty0 105169, 2022{\natexlab{b}}.

\bibitem[Roca(2023)]{10.52843/cassyni.6r5ry1}
L.~C. Roca.
\newblock {DNS-based optimisation of airfoils for Martian helicopters using
  PyFR}.
\newblock \emph{PyFR Seminar Series}, 2023.
\newblock \doi{10.52843/cassyni.6r5ry1}.

\bibitem[Roca(2021)]{10.52843/47ly7q}
L.~C. Roca.
\newblock {Martian Aerodynamics with PyFR}.
\newblock \emph{PyFR Seminar Series}, 2021.
\newblock \doi{10.52843/47ly7q}.

\bibitem[Bhaskaran and Stern(2024)]{doi:10.2514/6.2024-4127}
Rathakrishnan Bhaskaran and Eric~C. Stern.
\newblock \emph{Scale Resolving Simulations of Viking '75 Reentry Capsule Wake
  Flow}.
\newblock AIAA, 2024.
\newblock \doi{10.2514/6.2024-4127}.
\newblock URL \url{https://arc.aiaa.org/doi/abs/10.2514/6.2024-4127}.

\bibitem[Park(2021)]{10.52843/cassyni.536kkl}
J.~S. Park.
\newblock {High-Order Implicit Large-eddy Simulations of Flow around a
  Projectile using PyFR}.
\newblock \emph{PyFR Seminar Series}, 2021.
\newblock \doi{10.52843/cassyni.536kkl}.

\bibitem[Liang(2024)]{10.52843/cassyni.p3fyks}
T.~Liang.
\newblock {Actuator Line Model for Wind Turbine Wake Prediction with PyFR}.
\newblock \emph{PyFR Seminar Series}, 2024.
\newblock \doi{10.52843/cassyni.p3fyks}.

\bibitem[Liang and Hu(2024)]{LIANG2024121092}
Tianyang Liang and Changhong Hu.
\newblock Numerical simulation of wind turbine wake characteristics by flux
  reconstruction method.
\newblock \emph{Renewable Energy}, page 121092, 2024.
\newblock ISSN 0960-1481.
\newblock \doi{https://doi.org/10.1016/j.renene.2024.121092}.

\bibitem[Blanc(2023)]{10.52843/cassyni.dlsjz8}
N.~Blanc.
\newblock {Simulating a Thermoacoustic Engine With PyFR}.
\newblock \emph{PyFR Seminar Series}, 2023.
\newblock \doi{10.52843/cassyni.dlsjz8}.

\bibitem[Blanc et~al.(2024)Blanc, Laufer, Frankel, and Ramon]{BLANC2024122817}
Nathan Blanc, Michael Laufer, Steven Frankel, and Guy~Z. Ramon.
\newblock High-fidelity numerical simulations of a standing-wave thermoacoustic
  engine.
\newblock \emph{Applied Energy}, 360:\penalty0 122817, 2024.

\bibitem[Wan(2022)]{10.52843/cassyni.br5jn1}
Z.~Wan.
\newblock {Noise reduction of serrated trailing edges with implicit large eddy
  simulation using PyFR}.
\newblock \emph{PyFR Seminar Series}, 2022.
\newblock \doi{10.52843/cassyni.br5jn1}.

\bibitem[Yuan(2023)]{10.52843/cassyni.j59tff}
Z.~Yuan.
\newblock {Numerical simulations of aerofoil tonal noise reduction by roughness
  elements}.
\newblock \emph{PyFR Seminar Series}, 2023.
\newblock \doi{10.52843/cassyni.j59tff}.

\bibitem[Laufer(2021)]{10.52843/cassyni.nd09lk}
M.~Laufer.
\newblock {Implicit LES of NACA 0018 Airfoil with Active Flow Control}.
\newblock \emph{PyFR Seminar Series}, 2021.
\newblock \doi{10.52843/cassyni.nd09lk}.

\bibitem[Foysi(2023)]{10.52843/cassyni.2g7fx6}
H.~Foysi.
\newblock {Active control of compressible supersonic wall-bounded flow using
  direct numerical simulations with spanwise velocity modulation at the walls
  using PyFR}.
\newblock \emph{PyFR Seminar Series}, 2023.
\newblock \doi{10.52843/cassyni.2g7fx6}.

\bibitem[Cengiz(2023)]{10.52843/cassyni.hkqms2}
K.~Cengiz.
\newblock {Use of High-Order Curved Elements for Direct and Large Eddy
  Simulation of Flow over Rough Surfaces}.
\newblock \emph{PyFR Seminar Series}, 2023.
\newblock \doi{10.52843/cassyni.hkqms2}.

\bibitem[Regev(2023)]{10.52843/cassyni.5yklnq}
T.~Regev.
\newblock {GPU-Accelerated High-Fidelity Implicit LES of Coanda Cylinder Flow
  Instabilities}.
\newblock \emph{PyFR Seminar Series}, 2023.
\newblock \doi{10.52843/cassyni.5yklnq}.

\bibitem[Girayhan~Özbay(2022)]{h10.52843/cassyni.s1q5yf}
Ali Girayhan~Özbay.
\newblock {Unsteady 2D Flow Reconstruction around Arbitrary Shapes via
  Conformal Mapping aided Deep Neural Networks }.
\newblock \emph{PyFR Seminar Series}, 2022.
\newblock \doi{10.52843/cassyni.s1q5yf}.

\bibitem[Özbay and Laizet(2022)]{10.1063/5.0087488}
Ali~Girayhan Özbay and Sylvain Laizet.
\newblock {Deep learning fluid flow reconstruction around arbitrary
  two-dimensional objects from sparse sensors using conformal mappings}.
\newblock \emph{AIP Advances}, 12\penalty0 (4):\penalty0 045126, 04 2022.
\newblock ISSN 2158-3226.
\newblock \doi{10.1063/5.0087488}.
\newblock URL \url{https://doi.org/10.1063/5.0087488}.

\bibitem[Iyer et~al.(2019)Iyer, Witherden, Chernyshenko, and
  Vincent]{Iyer_Witherden_Chernyshenko_Vincent_2019}
A.~S. Iyer, F.~D. Witherden, S.~I. Chernyshenko, and P.~E. Vincent.
\newblock Identifying eigenmodes of averaged small-amplitude perturbations to
  turbulent channel flow.
\newblock \emph{Journal of Fluid Mechanics}, 875:\penalty0 758–780, 2019.
\newblock \doi{10.1017/jfm.2019.520}.

\bibitem[Aubry(2021)]{10.52843/cassyni.nqp2sp}
A.~Aubry.
\newblock {Gradient-Free Aerodynamic Shape Optimization using PyFR and MADS}.
\newblock \emph{PyFR Seminar Series}, 2021.
\newblock \doi{10.52843/cassyni.nqp2sp}.

\bibitem[Caros et~al.(2023)Caros, Buxton, and Vincent]{doi:10.2514/1.J063164}
Lidia Caros, Oliver Buxton, and Peter Vincent.
\newblock Optimization of triangular airfoils for martian helicopters using
  direct numerical simulations.
\newblock \emph{AIAA Journal}, 61\penalty0 (11):\penalty0 4935--4945, 2023.

\bibitem[Caros et~al.(2022)Caros, Buxton, Shigeta, Nagata, Nonomura, Asai, and
  Vincent]{doi:10.2514/1.J061454}
Lidia Caros, Oliver Buxton, Tsuyoshi Shigeta, Takayuki Nagata, Taku Nonomura,
  Keisuke Asai, and Peter Vincent.
\newblock Direct numerical simulation of flow over a triangular airfoil under
  martian conditions.
\newblock \emph{AIAA Journal}, 60\penalty0 (7):\penalty0 3961--3972, 2022.

\bibitem[Lusher and Sandham(2021)]{Lusher2021}
David~J. Lusher and Neil~D. Sandham.
\newblock Assessment of low-dissipative shock-capturing schemes for the
  compressible {T}aylor–{G}reen vortex.
\newblock \emph{AIAA Journal}, 59\penalty0 (2):\penalty0 533–545, February
  2021.
\newblock \doi{10.2514/1.j059672}.

\bibitem[Chapelier et~al.(2024)Chapelier, Lusher, Van~Noordt, Wenzel, Gibis,
  Mossier, Beck, Lodato, Brehm, Ruggeri, Scalo, and Sandham]{Chapelier2024}
Jean-Baptiste Chapelier, David~J. Lusher, William Van~Noordt, Christoph Wenzel,
  Tobias Gibis, Pascal Mossier, Andrea Beck, Guido Lodato, Christoph Brehm,
  Matteo Ruggeri, Carlo Scalo, and Neil Sandham.
\newblock Comparison of high-order numerical methodologies for the simulation
  of the supersonic {T}aylor–{G}reen vortex flow.
\newblock \emph{Physics of Fluids}, 36\penalty0 (5), May 2024.
\newblock \doi{10.1063/5.0206359}.

\bibitem[Sutherland(1893)]{Sutherland1893}
William Sutherland.
\newblock {LII}. {T}he viscosity of gases and molecular force.
\newblock \emph{The London, Edinburgh, and Dublin Philosophical Magazine and
  Journal of Science}, 36\penalty0 (223):\penalty0 507–531, December 1893.
\newblock \doi{10.1080/14786449308620508}.

\bibitem[Stefanin~Volpiani et~al.(2024)Stefanin~Volpiani, Chapelier,
  Schw\"{o}ppe, J\"{a}gersk\"{u}pper, and Champagneux]{StefaninVolpiani2024}
Pedro Stefanin~Volpiani, Jean-Baptiste Chapelier, Axel Schw\"{o}ppe, Jens
  J\"{a}gersk\"{u}pper, and Steeve Champagneux.
\newblock Aircraft simulations using the new {CFD} software from {ONERA},
  {DLR}, and {Airbus}.
\newblock \emph{Journal of Aircraft}, 61\penalty0 (3):\penalty0 857–869, May
  2024.
\newblock \doi{10.2514/1.c037506}.

\bibitem[Krais et~al.(2021)Krais, Beck, Bolemann, Frank, Flad, Gassner,
  Hindenlang, Hoffmann, Kuhn, Sonntag, and Munz]{Krais2021}
Nico Krais, Andrea Beck, Thomas Bolemann, Hannes Frank, David Flad, Gregor
  Gassner, Florian Hindenlang, Malte Hoffmann, Thomas Kuhn, Matthias Sonntag,
  and Claus-Dieter Munz.
\newblock {FLEXI}: A high order discontinuous {G}alerkin framework for
  hyperbolic–parabolic conservation laws.
\newblock \emph{Computers \& Mathematics with Applications}, 81:\penalty0
  186–219, January 2021.
\newblock \doi{10.1016/j.camwa.2020.05.004}.

\bibitem[Chapelier et~al.(2016)Chapelier, Lodato, and Jameson]{Chapelier2016}
J.-B. Chapelier, G.~Lodato, and A.~Jameson.
\newblock A study on the numerical dissipation of the spectral difference
  method for freely decaying and wall-bounded turbulence.
\newblock \emph{Computers \& Fluids}, 139:\penalty0 261–280, November 2016.
\newblock \doi{10.1016/j.compfluid.2016.03.006}.

\bibitem[Lusher et~al.(2021)Lusher, Jammy, and Sandham]{Lusher2021b}
David~J. Lusher, Satya~P. Jammy, and Neil~D. Sandham.
\newblock {OpenSBLI}: Automated code-generation for heterogeneous computing
  architectures applied to compressible fluid dynamics on structured grids.
\newblock \emph{Computer Physics Communications}, 267:\penalty0 108063, October
  2021.
\newblock \doi{10.1016/j.cpc.2021.108063}.

\bibitem[Vincent et~al.(2016)Vincent, Witherden, Vermeire, Park, and
  Iyer]{vincent2016towards}
P.~E. Vincent, F.~D. Witherden, B.~Vermeire, J.~S. Park, and A.~Iyer.
\newblock Towards green aviation with python at petascale.
\newblock In \emph{SC'16: Proceedings of the International Conference for High
  Performance Computing, Networking, Storage and Analysis}, pages 1--11. IEEE,
  2016.

\bibitem[{van Rees} et~al.(2011){van Rees}, Leonard, Pullin, and
  Koumoutsakos]{VANREES20112794}
Wim~M. {van Rees}, Anthony Leonard, D.I. Pullin, and Petros Koumoutsakos.
\newblock A comparison of vortex and pseudo-spectral methods for the simulation
  of periodic vortical flows at high reynolds numbers.
\newblock \emph{Journal of Computational Physics}, 230\penalty0 (8):\penalty0
  2794--2805, 2011.

\end{thebibliography}
